\documentclass[prx,twocolumn,nofootinbib,citeautoscript,longbibliography,notitlepage,superscriptaddress]{revtex4-1}

\usepackage[utf8]{inputenc}

\usepackage{csquotes}
\usepackage{graphicx}
\usepackage{dcolumn}
\usepackage{bm}
\usepackage{color}
\usepackage{amsmath}
\usepackage{tabularx}
\usepackage{epstopdf}
\usepackage{latexsym}
\usepackage{amssymb}
\usepackage{amsmath}
\usepackage{colortbl}
\usepackage{psfrag}
\usepackage{bbm}
\usepackage{bm}
\usepackage{titlesec}
\usepackage{dsfont}
\usepackage{feynmp}
\usepackage{slashed}
\usepackage{multirow}
\usepackage[tight]{subfigure}

\usepackage{comment}

\usepackage[papersize={8.5in,11in}]{geometry}
\usepackage{color}
\definecolor{darkblue}{rgb}{0.,0.,0.4}
\definecolor{darkred}{rgb}{0.5,0.,0.}
\definecolor{BlueViolet}{RGB}{138,43,226}
\definecolor{SkyBlue}{RGB}{30,144,255}
\definecolor{DarkGreen}{RGB}{0,100,0}
\usepackage[pdftex,colorlinks=true,linkcolor=darkblue,citecolor=blue,urlcolor=darkred]{hyperref}

\renewcommand{\epsilon}{\varepsilon}

\allowdisplaybreaks[3] 

\newcommand{\blue}[1]{\textcolor{blue}{#1}}

\usepackage{amsmath}
\usepackage{feynmp}
\usepackage{amssymb}
\usepackage{epsfig}
\usepackage{graphicx}
\usepackage{subfigure}
\usepackage{mathrsfs}
\usepackage{hyperref}
\usepackage{cleveref}
\usepackage{cases}
\usepackage{bbm}
\usepackage{xcolor}
\usepackage{tabularx}
\usepackage{array}
\usepackage{booktabs}
\usepackage{multirow}
\usepackage{lipsum}
\usepackage{float}
\usepackage{lmodern}
\usepackage{textcomp}
\usepackage{makecell}
\usepackage{comment}
\usepackage{threeparttable}

\allowdisplaybreaks[4]

\UseRawInputEncoding

\begin{document}

\title{Tunable phase transitions from semimetals to Chern insulators in two-dimensional quadratic-band-crossing materials}

\author{Wen-Hao Bian}
\affiliation{Department of Physics, Tianjin University, Tianjian 300072, P.R. China}
\affiliation{School of Physics, Nanjing University, Nanjing, Jiangsu 210093, P.R. China}

\author{Jing Wang}
\altaffiliation{Corresponding author: jing$\textunderscore$wang@tju.edu.cn}
\affiliation{Department of Physics, Tianjin University, Tianjian 300072, P.R. China}
\affiliation{Tianjin Key Laboratory of Low Dimensional Materials Physics and Preparing Technology,
Tianjin University, Tianjin 300072, P.R. China}

\date{\today}


\begin{abstract}

We systematically investigate how static symmetry-breaking perturbations and dynamic Floquet terms via a polarized light
manipulate the topological phase transitions in the two-dimensional quadratic-band-crossing-point (QBCP) materials.
The Berry curvature shows distinct behavior in such two situations. It is linearly and quadratically proportional
to the product of microstructural parameters $t_{x,z}$ for the former and the latter, respectively.
The static perturbation eliminates the QBCP and opens an energy gap, which leads to the momentum-inversion symmetry of
Berry curvature. This yields a nontrivial Chern number determined by the microstructural
parameters. In contrast, we demonstrate that either a circularly or an elliptically polarized light
breaks the time-reversal symmetry, transforming the QBCP semimetal into a Chern insulator with a quantized anomalous Hall
conductivity $\sigma_{xy} = Ce^2/\hbar$, where the Chern number is governed by the polarization angle.
Moreover, the linear polarization preserves the central antisymmetry of the Berry curvature, giving rise to a topologically trivial insulator, and the potential optical signatures to probe distinct topological transitions are also discussed.
These results establish a tunable topological phase transition from a QBCP semimetal to Chern insulator
in the two-dimensional QBCP materials.
\end{abstract}


\maketitle

\section{Introduction}

Topological insulators (TIs) are one of the most important and hottest topics in contemporary condensed matter
physics, which fundamentally differ from conventional band insulators by exhibiting both insulating bulk states and
topologically protected metallic edge/surface states~\cite{Haldane1988PRL,Kane2005PRL,Kane2010RMP,Xiao2010RMP,Zhang2011RMP}. These unique
properties have been attracting a plethora of both experimental and theoretical studies~\cite{Kane2005PRL,Kane2005PRL,Xiao2010RMP,Kane2010RMP,Zhang2011RMP,Chiu2016RMP,Feng2012SCPMA,Yan2012RPP,Ando2013JPSJ,
Bardarson2013RPP,Beenakker2013ARCMP,Zhang2013RSSRRL,Bansil2016RMP,Lapano2020PRM}.
Whether the time-reversal symmetry (TRS) is preserved or broken gives rise to two distinct categories of TIs in two dimensions.
One is the $\mathbb{Z}_2$ topological insulator protected by the TRS and characterized the $\mathbb{Z}_2$ index~\cite{Kane2005PRL,Bernevig2006SCI}.
On the contrary, the other corresponds to the Chern insulator that breaks the TRS and is characterized by a nontrivial
Chern number and nonzero anomalous Hall conductivity (AHC)~\cite{Thouless1982PRL,Haldane1988PRL}.
Particularly, Chern insulators have become a crucial research frontier in topological quantum materials.
Beyond the Haldane model~\cite{Haldane1988PRL}, both the higher-order Chern insulators~\cite{Chern2020PRL,Chen2021PRX}
and dual Chern insulators have been proposed~\cite{Bai2025ACS}. Besides, several groups~\cite{Haldane1988PRL,Thouless1982PRL,Chern2020PRL,Chen2021PRX,Bai2025ACS,Lu2025PRL,
Wang2025arXiv,He2025arXiv,Emanuel2025arXiv,Lin2025arXiv,Chang2013SCI,Xiang2023NatCommun,Huang2024JPCM} recently
advocate the fractional Chern insulators which host fractionalized excitations. Since the quantum anomalous Hall
effect in Chern insulator was observed in $\mathrm{Cr}_{0.15}(\mathrm{Bi}_{0.1}\mathrm{Sb}_{0.9})_{1.85}\mathrm{Te}_3$~\cite{Chang2013SCI},
the experimental studies on Chern insulators also made significant progress including quantum computing~\cite{Xiang2023NatCommun},
topological state manipulation~\cite{Xiang2023NatCommun,Huang2024JPCM}, and quantum simulation~\cite{Ji2024Nature,Lei2024CPL,Zhang2023PRB}.

Particularly, topological phase transitions have garnered extensive studies in various kinds
of semimetals, including Dirac semimetals characterized by linear dispersions~\cite{Castro2009RMP,Lu2010PRB,Bernevig2006SCI,
Shen2011SPIN,Taguchi2020PRB,Vargiamidis2022arXiv,Lu2022PRB,Wang2023PRB,Mo2024arXiv}
and semi-Dirac materials equipped with hybrid linear-quadratic dispersions~\cite{Zhang2005Nature,Huang2015PRB,Mondal2022PRB,Chen2018PRB,Saha2016arXiv}.
In recent years, attention has gradually shifted to a new type of semimetal. Considering this type of semimetal, upper and lower bands
parabolically touching at a quadratic band crossing point (QBCP)~\cite{Chong2008PRB,Sun2008PRB,
Sun2009PRL,Vafek2014PRB,Yao2022PRR,Wang2024AP,Mandal2019CMP,Bera2021JPCM,Lu2024PRB}. The quadratic dispersion in such systems leads to an enhanced density of states and gives rise to unique symmetry-protected topological properties that are absent in their linear counterparts~\cite{Chong2008PRB,Sun2009PRL,Vafek2014PRB,Liquito2024PRB,Jung2023arXiv,Wan2023PRL,Ji2022PRB,Wu2022SCPMA,Sobrosa2024PRB}. Additionally, certain interaction-driven topological phase transitions in these materials have been reported~\cite{Liquito2024PRB,Wan2023PRL,Ji2022PRB,Sobrosa2024PRB,Wang2017PRB,DZZW2020PRB,Janssen2018PRB}.
While the perturbation control of topological phases was advocated in Dirac
systems~\cite{Saha2016arXiv,Mo2024arXiv,Pan2015SR},
its application to the QBCP semimetals, however, has not yet been sufficiently investigated.
Particularly, there exist two important issues to be delved into.
Initially, it is worth examining how static and dynamic perturbations induced by tunable external optical fields
distinctively influence the topological properties of QBCP systems. Besides, it is imperative to explore the
potential of these external optical fields serve as a suitable tool to control the topological phase transitions
between trivial and nontrivial phases.

To this end, we systematically investigate the effects of static perturbations and tunable
optical fields as well as their interplay on the topological properties of 2D QBCP semimetals.
Specifically, we begin with the study of a static perturbation.
It opens an energy gap with breaking TRS and causes the Berry curvature
symmetric under momentum inversion, yielding Chern insulators with nontrivial quantized Chern numbers
$C = \pm\mathrm{sgn}(t_x t_z)$ governed by structural parameters. Next, a polarized light irradiation is imposed
on the QBCP system by adopting the Floquet theory~\cite{Oka2009PRB,Rudner2020NRP,Liu2025arXiv,Fu2025arXiv,Yokoyama2025arXiv}.
We find that either a circularly or an elliptically tunable polarized light breaks the TRS, transforming the QBCP semimetal
into a Chern insulator with $C=\pm\mathrm{sgn}(\phi)$. In comparison, the linear polarization preserves trivial topology.
Hereby, the polarization angle ($\phi$) acts as a tunable parameter
to switch a trivial and Chern insulator. Moreover, we briefly examine the competition between
static and dynamically optical perturbations, which measured by the parameters $m$ and $A_0$, respectively.
At $m \gg A_0$, the former dominates with $C = \pm\mathrm{sgn}(t_x t_z)$, whereas $A_0 \gg m$
enables the latter become dominant. Furthermore, the behavior of AHC accompanied by the topological phase transitions
at zero temperature is addressed. Both types of perturbations exhibit a universal quantization
$\sigma_{xy} = Ce^2/\hbar$ with $C = 1$ at the Fermi energy but instead present non-universal scalings
in other regimes that are modulated by the parameters $t_I$ and $A_0/\phi$ for static and dynamic
perturbations, respectively.
At last, we briefly review the intrinsic Euler topology of the 2D QBCP semimetal and
then evaluate the feasibility of using optical probes, such as circular dichroism~\cite{Jankowski2025PRB, Chau2025PRB,Tran2017SA,Asteria2019NP},
optical-weight measurements~\cite{Jankowski2025PRB, Chau2025PRB}, and higher-order nonlinear photoconductivity~\cite{Jankowski2025PRB, Chau2025PRB,Tran2017SA,Asteria2019NP},
to distinguish between different topological phases. Together, these discussions offer
a theoretical framework and distinct experimental signatures for potential
topological phase transitions.
In contrast to earlier studies on Dirac-system transitions and the non-Abelian properties of QBC semimetals~\cite{Lu2010PRB,Bernevig2006SCI,
Shen2011SPIN,Taguchi2020PRB,Vargiamidis2022arXiv,Lu2022PRB,Wang2023PRB,Mo2024arXiv,Jankowski2025PRB},
this work provides a systematic investigation of topological phase transitions in 2D QBCP semimetals
driven by both static fields and Floquet dynamics.  These results resolve the interplay between symmetry constraints, geometric
Berry response, and topological phase transitions in QBCP materials, providing theoretical references for
further studies of the related quantum materials.

The rest of this paper is organized as follows. In Sec.~\ref{Sec_model}, we present three
effective models, including the free QBCP model, static-perturbation scenario, and Floquet-driven scenario, to construct the effective Hamiltonian.  Subsequently, Sec.~\ref{Sec_Berry_Chern} is followed to compute the Berry curvature and Chern number, and perform
a detailed analysis of the topological phase transitions as well as provide the overall phase diagrams.
Thereafter, we within Sec.~\ref{Sec_Hall_conductivity} calculate the anomalous Hall conductivity
arising from TRSB and nontrivial Berry curvature.
Moreover, Sec.~\ref{Sec_optical} provides the related non-Abelian topological properties and underlying
experimental discusses on the optical properties. Finally, Sec.~\ref{Sec_summary} concludes with
a brief summary of the key results.

\section{Effective models}\label{Sec_model}

To systematically investigate the topological properties of 2D QBCP semimetals,
we within this work consider three distinct cases, which are denominated as
the free model that preserves both $C_4$ point-group symmetry and TRS~\cite{Sun2008PRB,Sun2009PRL,Vafek2014PRB,Yao2022PRR},
the perturbative scenario incorporating an external field to break TRS~\cite{Saha2016arXiv,Chong2008PRB}, and the
Floquet scenario with a time-periodic driving~\cite{Saha2016arXiv,Polkovnikov2014arXiv,Benhaida2024arXiv,Kibis2024arXiv,Asgari2025arXiv,
Oka2009PRB,Rudner2020NRP,Liu2025arXiv,Fu2025arXiv,Yokoyama2025arXiv}, respectively.

\subsection{Free model}\label{Sec_free_model}

\begin{figure}[htpb]
\centering
\includegraphics[width=3in]{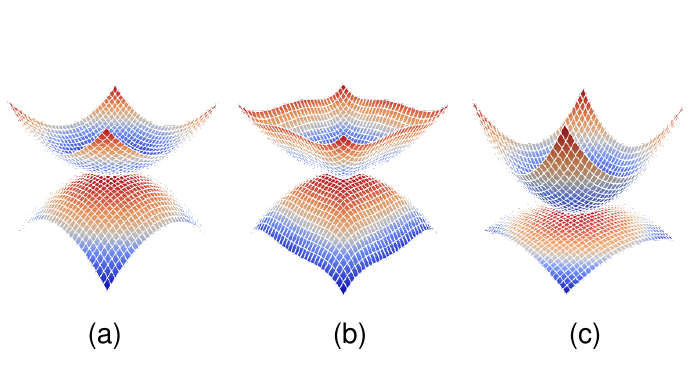}
\vspace{-0.5cm}
\caption{(Color online) Schematic dispersions for the 2D QBCP semimtals:  (a) $t_x= t_z$,
(b) $t_x\neq t_z$ with $t_I=0$, and (c) $t_x=t_z$ with $t_I\neq0$.}
\label{dispersion}
\end{figure}

At first, let us introduce the free model.
The free effective continuous Hamiltonian for 2D QBCP semimetals with a checkerboard lattice is given by~\cite{Sun2008PRB,Sun2009PRL,Vafek2014PRB,Yao2022PRR}
\begin{equation}
H_0=\sum_{\sigma=\uparrow\downarrow}\sum_{\bm{k}<|\Lambda|}
\Psi_{\bm{k}\sigma}^\dagger \mathcal{H}_0(\bm{k})\Psi_{\bm{k}\sigma},\label{H_0}
\end{equation}
where $\Lambda$ serves as the momentum cutoff associated with the lattice constant, and $\Psi_{\bm{k}\sigma} = (c_{a\sigma}, c_{b\sigma})^\mathrm{T}$ represents the sublattice spinor with the indexes $a$ and $b$
denoting sublattices and $\sigma$ being the spin of electron. The Hamiltonian density takes the form:
\begin{equation}
\mathcal{H}_0(\bm{k}) = t_I \bm{k}^2 \sigma_0 + 2 t_x k_x k_y \sigma_1 + t_z (k_x^2 - k_y^2) \sigma_3.\label{H_0_2}
\end{equation}
Hereby, $t_I$, $t_x$, and $t_z$ correspond to material-dependent parameters and the Pauli matrices $\sigma_{1,2,3}$  with
the identity matrix $\sigma_0$ act on the sublattice space. The Hamiltonian density can also be written compactly as
\begin{eqnarray}
\mathcal{H}_0=t_I\bm{k}^2\sigma_0+\bm{d}_0(\bm{k})\cdot\bm{\sigma},\label{H_01}
\end{eqnarray}
with $\bm{d}_0(\bm{k})\equiv\left(2t_xk_xk_y,~0,~t_z(k_x^2-k_y^2)\right)$ and $\bm{\sigma}\equiv(\sigma_1,\sigma_2,\sigma_3)$
is the Pauli-matrix vector.

Diagonalizing $\mathcal{H}_0(\bm{k})$ yields the parabolical energy eigenvalues ~\cite{Sun2008PRB,Sun2009PRL,Vafek2014PRB,Yao2022PRR,Wang2024AP,Wang2021NPB},
\begin{equation}
E^\pm_{\bm{k}}\!=\!\frac{\bm{k}^2\left(\lambda\pm \sqrt{\cos^2\eta\cos^22\theta_{\bm{k}}+\sin^2\eta\sin^22\theta_{\bm{k}}}\right)}{\sqrt{2}m},
\end{equation}
where
\begin{eqnarray}
m&=&\frac{1}{\sqrt{2(t_x^2+t_z^2)}},
\lambda=\frac{t_I}{\sqrt{t_x^2+t_z^2}},\\
\cos\eta&=&\frac{t_z}{\sqrt{t_x^2+t_z^2}},
\sin\eta=\frac{t_x}{\sqrt{t_x^2+t_z^2}},\\
\cos\theta_{\bm{k}}&=&\frac{k_x}{\sqrt{k_x^2+k_y^2}},
\sin\theta_{\bm{k}}=\frac{k_y}{\sqrt{k_x^2+k_y^2}},
\end{eqnarray}
with $\theta_{\bm{k}}$ specifying the direction of momentum.
This energy dispersion features two bands: an upper band and a lower band, which touch quadratically
at a quadratic band crossing point (QBCP) with $\bm{k}=0$ when the condition $|t_I| < \mathrm{min}(|t_x|, |t_z|)$
is satisfied. It is of particular importance to highlight that the particle-hole (PH) and
rotational symmetries depend upon the microstructural parameters as schematically depicted
in Fig.~\ref{dispersion}. A nonzero parameter $t_I$ explicitly breaks the PH symmetry,
and the rotational symmetry is preserved when $t_x = t_z$ but broken otherwise.

In particular, the 2D QBCP free Hamiltonian~(\ref{H_0_2}) exhibits two fundamental symmetries. Specifically,
it satisfies the time-reversal symmetry (TRS) under $\mathcal{T}\mathcal{H}_0(\bm{k}) \mathcal{T}^{-1}=\mathcal{H}_0(-\bm{k})$
due to a paucity of $\sigma_2$ term~\cite{Sun2009PRL, Vafek2014PRB}.
Besides, the Hamiltonian respects the spatial inversion symmetry expressed as
$\mathcal{P} \mathcal{H}_0(\bm{k}) \mathcal{P}^{-1}=\mathcal{H}_0(-\bm{k})$.
Herein, $\mathcal{T}$ and $\mathcal{P}$ are the time-reversal and parity operators,
respectively.

The $\sigma_1$ and $\sigma_3$ terms in the free Hamiltonian~(\ref{H_0_2}) originate from the symmetries of
the $d_{xy}$- and $d_{x^2-y^2}$-orbital wavefunctions. These terms give rise to a net Berry flux of $\pm 2\pi$
around the QBCP~\cite{Sun2009PRL, Vafek2014PRB}. This accordingly indicates that the free Hamiltonian~(\ref{H_0_2})
describes a robust semimetal state that hosts a QBCP at $\bm{k}=0$ and is protected by TRS.
A critical question then arises: Can the system exhibit topologically nontrivial states if the QBCP is removed or if TRS is explicitly broken? To address this, we propose two strategies for realizing time-reversal symmetry breaking (TRSB), which will be detailed in Sec.~\ref{subSec_TRSB_model}.


\subsection{Perturbative and Floquet scenarios}\label{subSec_TRSB_model}

Next, we bring out the other two scenarios. To enable certain topological phase transition
and induce nontrivial topological properties in the 2D QBCP system,
it is essential to breaking TRS~\cite{Kane2010RMP,Xiao2010RMP,Zhang2011RMP,Chiu2016RMP}. To this end, we bring out two
distinct approaches: (a) introducing a generic perturbation
term to the free model~(\ref{H_0_2}), and (b) employing the light-matter interaction as a realistic
symmetry-breaking mechanism.

\subsubsection{\blue{Perturbation scenario}}\label{subsubSec_toy_model}

As a simple toy scenario and control case, let us begin with removing the QBCP and opening an energy gap
via the introduction of a perturbation constant term $\delta\mathcal{H}$, which can be realized by
interactions, impurities, periodic potentials, etc.~\cite{Zhou2007NM,Bao2021PRL,Wang2025PRB,Wang2013PRL,Saha2016arXiv,Bansil2016RMP}.
Then, the free model~(\ref{H_0_2}) is casted into~\cite{Saha2016arXiv,Wang2013PRL,Huang2015PRB}
\begin{eqnarray}
\mathcal{H}_{\mathrm{pert}}(\bm{k})=\mathcal{H}_0(\bm{k})+\delta\mathcal{H},
\end{eqnarray}
we highlight that the perturbation term $\delta\mathcal{H}$ is not arbitrarily chosen~\cite{Wu2016PRL,Chong2008PRB}.

\begin{figure}[htpb]
\centering
\subfigure[]{\includegraphics[width=1in]{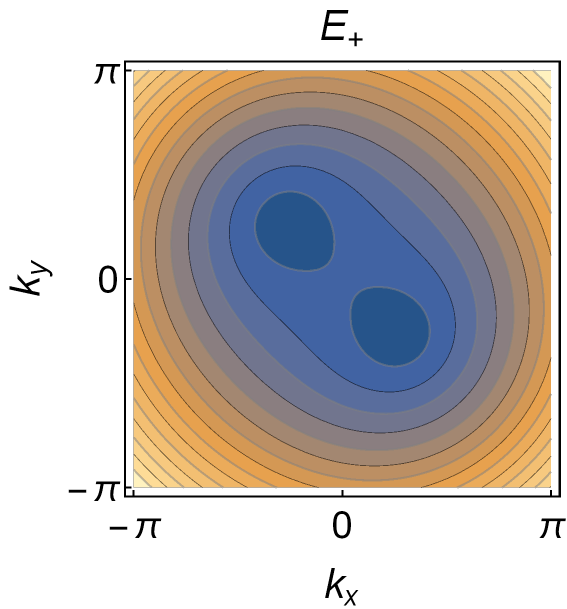}}
\subfigure[]{\includegraphics[width=1in]{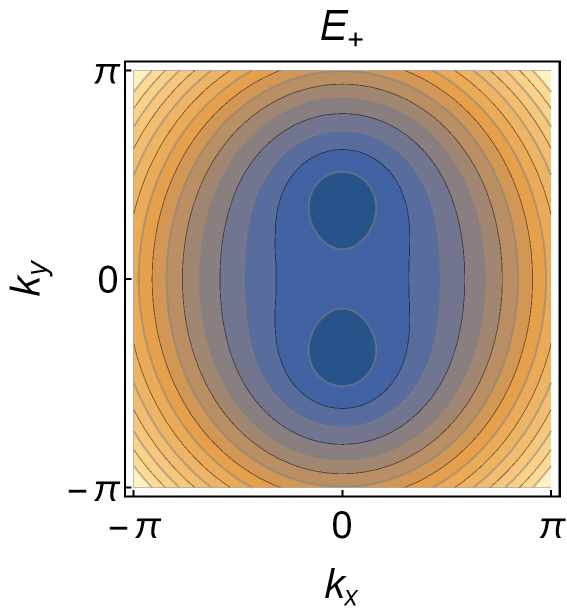}}
\subfigure[]{\includegraphics[width=1in]{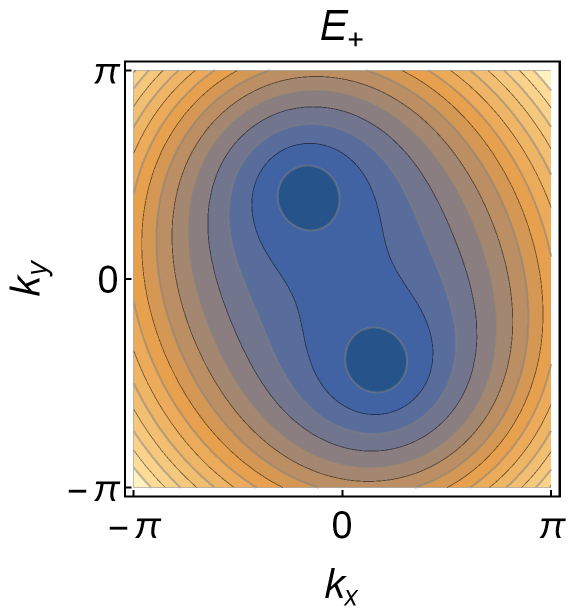}}\\
\vspace{-0.2cm}
\caption{(Color online) Splitting of the quadratic touching points under the perturbative term $\delta\mathcal{H}=m_i\sigma_i$
with $i=1,3$: (a) $m_1\neq0, m_3=0$, (b) $m_3\neq0,m_1=0$, and (c) $m_1=m_3\neq 0$.}
\label{dispersion_per}
\end{figure}

To break the TRS of the free QBCP system, we introduce an external symmetry-breaking term. This term is treated as a
finite, tunable parameter (not restricted to small values) and serves as an effective gap-opening field,
following the approach established in Refs.~\cite{Saha2016arXiv,Wang2013PRL}.
Specifically, when we consider $\delta\mathcal{H}=m_i\sigma_i$ with $i=1,3$, the QBCP cannot be removed but splitting into
two distinct quadratic touching points as shown in Fig.~\ref{dispersion_per}.
For $m_1 \neq 0, \, m_3 = 0$, the QBCP splits into two points at $\bm{k} = \sqrt{m_1/2t_x}(\pm1,\mp1)$ ($m_1t_x > 0$) or $\bm{k} = \sqrt{-m_1/2t_x}(\pm1,\pm1)$ ($m_1t_x < 0$), aligned with $k_x = \pm k_y$.
For $m_1 = 0, \, m_3 \neq 0$, the points shift to $k_y = 0$ ($m_3 > 0$) or $k_x = 0$ ($m_3 < 0$). In the general
case ($m_1, m_3 \neq 0$), the points locate at $\bm{k} = (\pm\sqrt{\alpha/2}, \mp m_1/(t_x\sqrt{2\alpha}))$ with
$\alpha = -m_3/t_z + \sqrt{(m_3/t_z)^2 + (m_1/t_x)^2}$~\cite{Chong2008PRB}.

In consequence, we are left with $\delta\mathcal{H}=m\sigma_2$ with $m>0$ as a symmetry-breaking perturbation.
The total Hamiltonian dubbed the perturbative scenario becomes
\begin{eqnarray}
\mathcal{H}_{\mathrm{pert}}(\bm{k})=t_I\bm{k}^2\sigma_0+\bm{h}(\bm{k})\cdot\bm{\sigma},\label{eq9}
\end{eqnarray}
with $\bm{h}(\bm{k})=(2t_x k_x k_y, m, t_z(k_x^2-k_y^2))$.
The $\sigma_2$ perturbation term induces critical effects.
The QBCP is eliminated and a direct energy gap $\Delta = 2m$ emerges at $\bm{k} = \bm{0}$.
More importantly, the TRS is explicitly broken, namely $\mathcal{T}\mathcal{H}_{\mathrm{per}}(\bm{k})\mathcal{T}^{-1}\neq \mathcal{H}_{\mathrm{per}}(-\bm{k})$, which are expected to generates nontrivial Berry curvature and nontrivial topological
properties~\cite{Saha2016arXiv,Kane2010RMP,Xiao2010RMP,Zhang2011RMP,Chiu2016RMP}.

\vspace{0.3cm}

\subsubsection{Floquet scenario}\label{Subsubsec_Floquet}

In order to bridge theoretical scenario with experiment, we go beyond the
toy scenario presented in Sec.~\ref{subsubSec_toy_model} and employ light-matter interaction as a realistic symmetry-breaking mechanism,
i.e, the Floquet theory~\cite{Polkovnikov2014arXiv,Saha2016arXiv,Benhaida2024arXiv,Kibis2024arXiv,Asgari2025arXiv,
Oka2009PRB,Rudner2020NRP,Liu2025arXiv,Fu2025arXiv,Yokoyama2025arXiv}.

\begin{figure*}[htpb]
\centering
\subfigure[]{\includegraphics[width=2.3in]{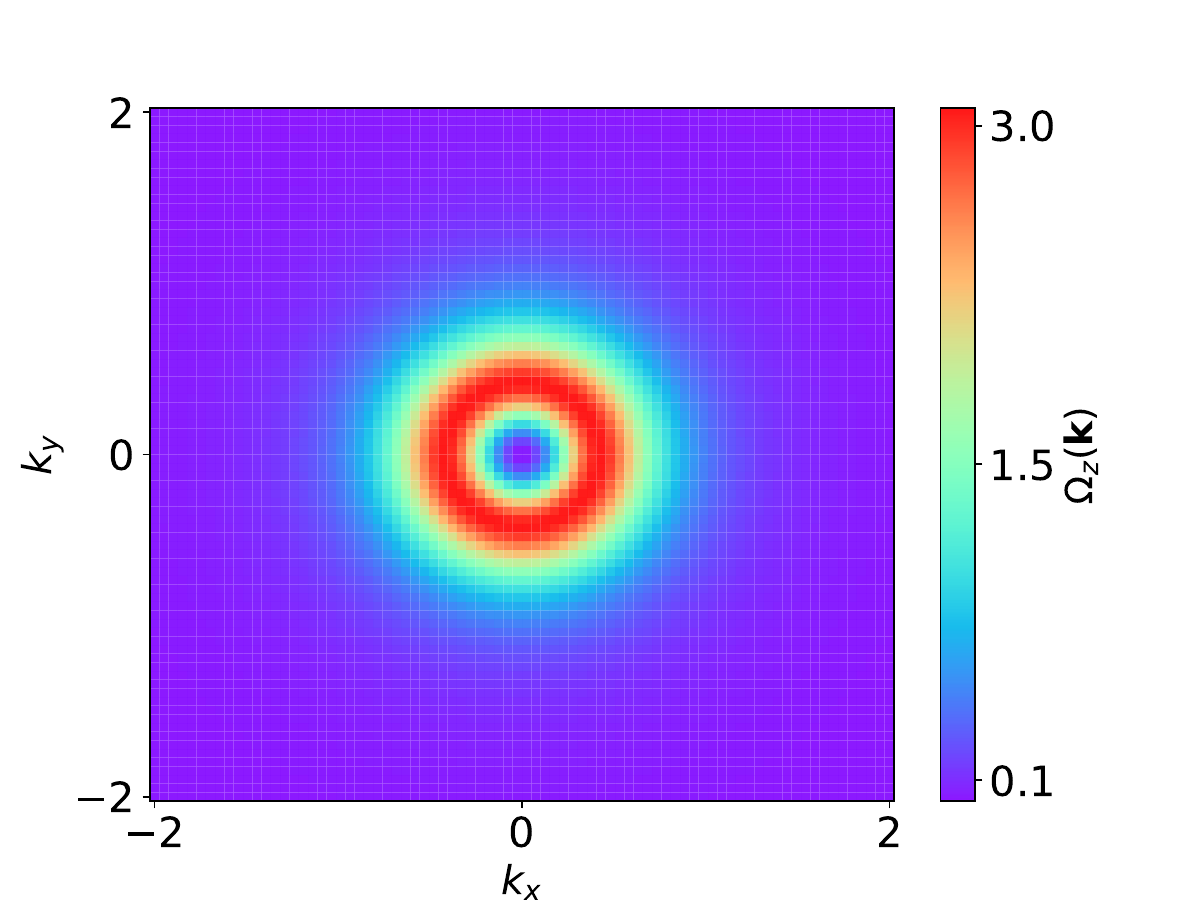}}
\hspace{-0.5cm}
\subfigure[]{\includegraphics[width=2.3in]{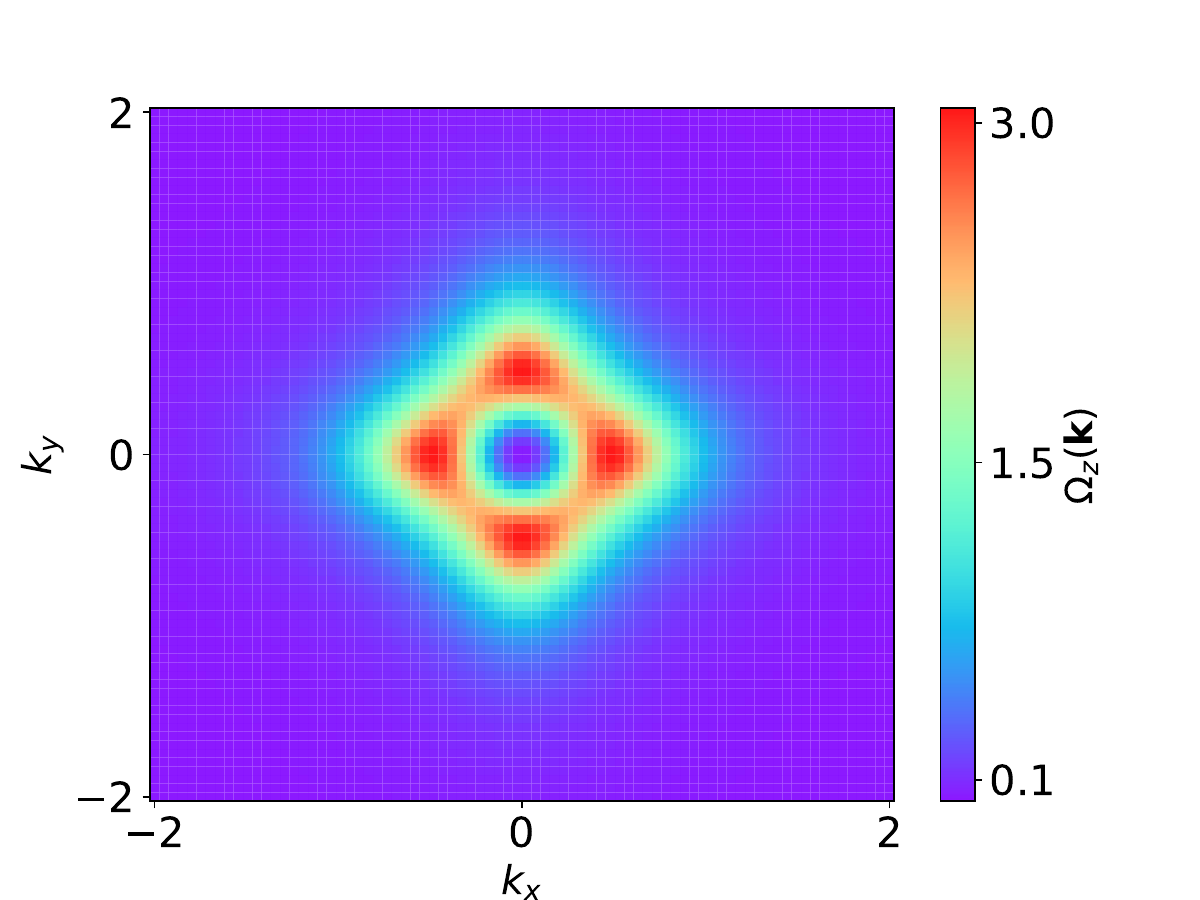}}
\hspace{-0.5cm}
\subfigure[]{\includegraphics[width=2.3in]{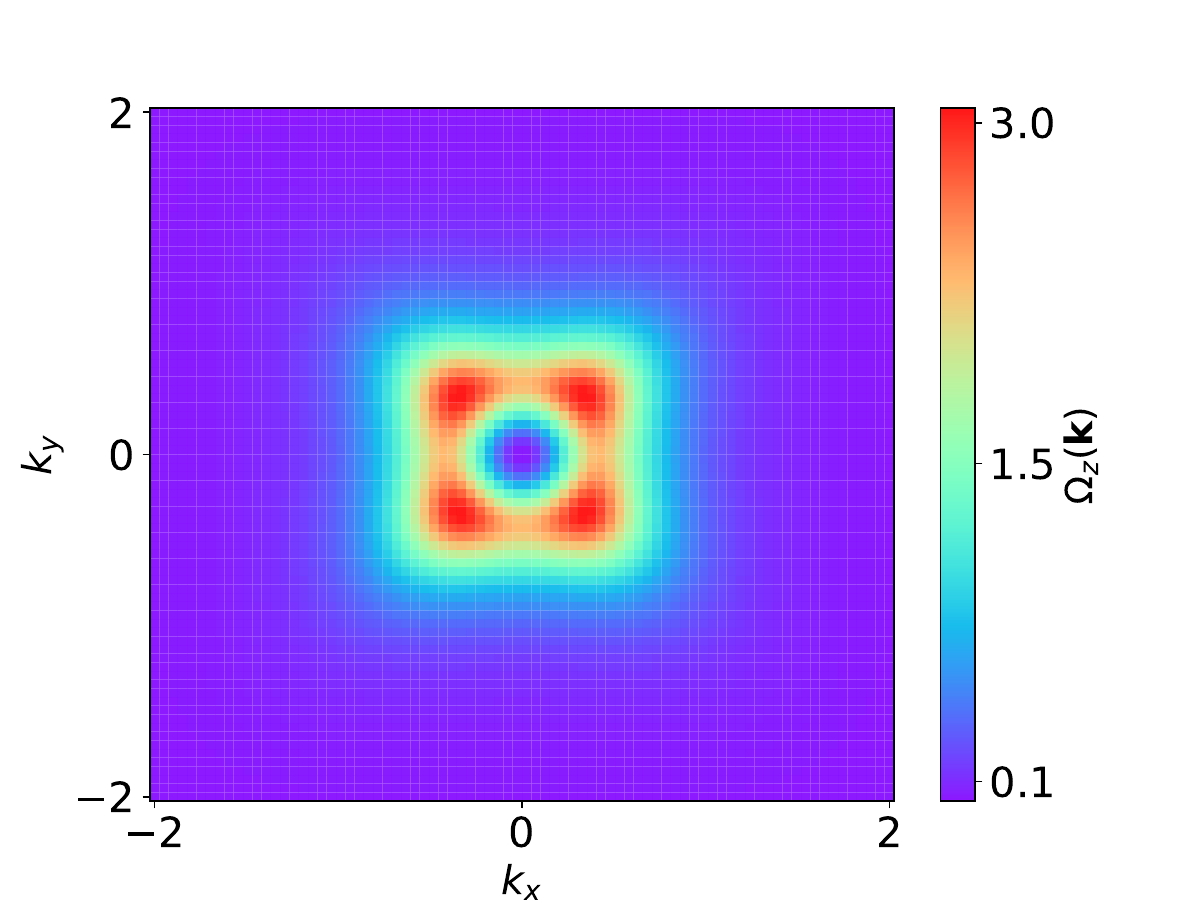}}
\vspace{-0.2cm}
\caption{(Color online) Momentum dependence of Berry curvature for the perturbation scenario:
(a) $t_x=t_z=1.6$, (b) $t_x=1.6,~t_z=1.2$, (c) $t_x=1.2,~t_z=1.6$ with $m=0.4$.}
\label{bc_typeI_pert}
\end{figure*}

Within the Floquet framework, circularly polarized light irradiation provides a controlled pathway to dynamically break TRS without static lattice alterations~\cite{Saha2016arXiv,Polkovnikov2014arXiv,Benhaida2024arXiv,Kibis2024arXiv,Asgari2025arXiv,
Oka2009PRB,Rudner2020NRP,Liu2025arXiv,Fu2025arXiv,Yokoyama2025arXiv,Ma2024arXiv,Mola2025arXiv,Li2025arXiv}.
Specifically, the time-periodic modulation is implemented by coupling the Hamiltonian of free model~(\ref{H_0_2}) to a monochromatic radiation field $\bm{A}(t) = (A_x, A_y)=A_0(\sin(\omega t),~\sin(\omega t+\phi))$ where the spatial dependence is neglected as the wavelength of the field is much larger compared to the sample size. Hereby, $\omega$ and $A_0$ denote the frequency and amplitude of the driving field, respectively.
The polarization state of light is parameterized by the phase angle $\phi \in (-\pi, \pi]$, which determines the relative phase between orthogonal components of the electric field. This yields three distinct polarization types:
the circular polarization (CPL) at $\phi = \pm\pi/2$,  the linear polarization (LPL)
at $\phi = 0,\pi$, and the elliptical polarization (EPL) for all other values of $\phi$.
Performing the minimum coupling via $\bm{k} \to \bm{k}+e\bm{A}(t)$ with $e$ being the electric charge gives rise to the
new Hamiltonian~\cite{Saha2016arXiv},
\begin{eqnarray}
\mathcal{H}(\bm{k},t)=\mathcal{H}_0(\bm{k})+H_1\sigma_0+H_2\sigma_1+H_3\sigma_3\label{eq6}
\end{eqnarray}
where
\begin{widetext}
\begin{eqnarray}
H_1&=&t_I e A_0\big\{2k_x\sin(\omega t)+eA_0\sin^2(\omega t)+\sin(\omega t+\phi)\big[2k_y+eA_0\sin(\omega t+\phi)\big]\big\},\\
H_2&=&2 t_x e A_0\big\{k_x\sin(\omega t+\phi)+\sin(\omega t)\big[k_y+e A_0\sin(\omega t+\phi)\big]\big\},\\
H_3&=&\frac{t_z eA_0}{2}\big\{-eA_0\cos(2\omega t)+eA_0\cos(2\omega t+2\phi)+4k_x\sin(\omega t)-4k_y\sin(\omega t+\phi)\big\}.
\end{eqnarray}
\end{widetext}
In the spirt of Floquet theory~\cite{Saha2016arXiv,Polkovnikov2014arXiv,Benhaida2024arXiv,Kibis2024arXiv,Asgari2025arXiv,
Oka2009PRB,Rudner2020NRP,Liu2025arXiv,Fu2025arXiv,Yokoyama2025arXiv}, the stroboscopic dynamics of
the time-periodic Hamiltonian can be mapped to an effective static
Floquet Hamiltonian $\mathcal{H}_\mathrm{F}(\bm{k})$. This Floquet Hamiltonian formally satisfies~\cite{Saha2016arXiv,Polkovnikov2014arXiv,Benhaida2024arXiv,Kibis2024arXiv,Asgari2025arXiv,
Oka2009PRB,Rudner2020NRP,Liu2025arXiv,Fu2025arXiv,Yokoyama2025arXiv}
\begin{eqnarray}
\mathcal{H}_{\mathrm{F}}(\bm{k})=\frac{i}{\hbar T}\ln\big[\mathrm{U}(t_0+T,t_0)\big]\label{eq8},
\end{eqnarray}
where $\mathrm{U}(t,t_0)=\mathcal{T}_t\exp\left[-\frac{i}{\hbar
}\int_{t_0}^tdt'~\mathcal{H}(\bm{k},t')\right]$ is
the  evolution operator with $\mathcal{T}_t$ denoting the time-ordering operator and $t_0$ the initial time of the perturbation~\cite{Polkovnikov2014arXiv,Saha2016arXiv,Benhaida2024arXiv,Kibis2024arXiv}.
After several analytical calculations in Appendix.~\ref{AppB}, we finally arrive at the effective Hamiltonian
of the Floquet scenario
\begin{eqnarray}
\mathcal{H}_{\mathrm{F}}(\bm{k})=t_I(e^2A_0^2+\bm{k}^2)\sigma_0+\bm{d}(\bm{k})\cdot\bm{\sigma}\label{eq16},
\end{eqnarray}
where the vector takes the form of
\begin{equation}
\bm{d}(\bm{k})=(t_xe^2A_0^2\cos\phi+2t_xk_xk_y,m_{\mathrm{eff}},t_z(k_x^2-k_y^2)),\label{eq21}
\end{equation}
with the effective mass term depending on the momentum $\bm{k}$ and the strength of field $\mathbf{A}$,
\begin{widetext}
\begin{eqnarray}
m_{\mathrm{eff}}&=&\frac{eA_0t_xt_z}{4\omega\hbar}\bigg\{8e^2A_0^2\big[1+\cos(2\phi)\big]k_y
+20eA_0\sin\phi~k_x^2+12eA_0\sin\phi~k_y^2+4eA_0\sin(2\phi)k_xk_y\nonumber\\
&&-16 e^2A_0^2\cos\phi~k_x+16(\cos\phi~k_x-k_y)\bm{k}^2+e^3 A_0^3\big[3\sin\phi+\sin(3\phi)\big]\bigg\}.
\end{eqnarray}
\end{widetext}
The Floquet Hamiltonian in Eq.~(\ref{eq16}) induces a band gap analogous to the static perturbation scenario in Sec.~\ref{subsubSec_toy_model}. This gap $\Delta$ emerges near $\bm{k} \to \bm{0}$ and depends on $t_x$, $t_z$, driving amplitude $A_0$, polarization angle $\phi$, and frequency $\omega$:
\begin{eqnarray}
\Delta = \sqrt{\left(\delta_1
\cos\phi\right)^2+\frac{1}{4}\left[\delta_2\left(3\sin\phi+\sin(3\phi)\right)\right]^2},
\end{eqnarray}
where $\phi$ plays a crucial role in the magnitude of gap,
and the $\delta_1=2t_xe^2A_0^2$ and $\delta_2= \frac{e^4A_0^4 t_xt_z}{\hbar\omega}$
correspond to the gap for LPL ($\phi=0,~\pi$) and CPL ($\phi=\pm \pi/2$), respectively.

On the basis of these scenarios, we are going to address the questions raised
in the end of Sec.~\ref{Sec_free_model}
via carefully studying the topological nontrivial properties
for both perturbative and Floquet scenarios. The resulting topological properties--including Berry curvature,
Chern numbers, and anomalous Hall conductivity--will be systematically analyzed in Sec.~\ref{Sec_Berry_Chern} and Sec.~\ref{Sec_Hall_conductivity}.


\begin{figure}[htpb]
\centering
\subfigure[]
{\includegraphics[width=1.5in]{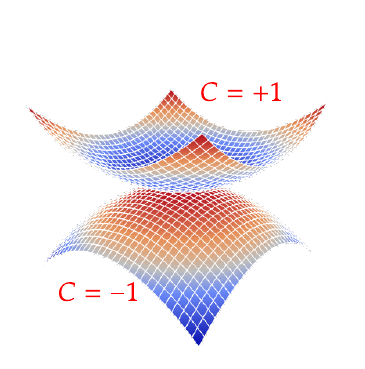}}
\subfigure[]
{\includegraphics[width=1.5in]{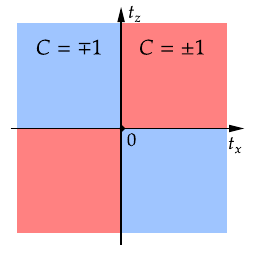}}
\vspace{-0.2cm}
\caption{(Color online) Schematic descriptions of the basic results
for the perturbative scenario: (a) nontrivial Chern numbers of the upper and
lower bands and (b) $t_x,t_z$-tuned topological phase transitions.}
\label{Chern}
\end{figure}

\section{Berry curvature and Chern number}\label{Sec_Berry_Chern}

Within this section, we systematically investigate the Berry curvature, Chern number,
and potential topological phase transitions for the effective models established in Sec.~\ref{Sec_model}.
A generic two-level Hamiltonian can be comprehensively described by~\cite{Xiao2010RMP,Zhang2011RMP,Kane2010RMP}
$\mathcal{H}_0(\bm{k}) = d_0(\bm{k})\sigma_0 + \bm{d}(\bm{k})\cdot\bm{\sigma}$
where $d_0(\bm{k})$ represents a scalar function of momentum, while $ \bm{d}(\bm{k})$
constitutes a momentum-dependent vector parameter. The former can induce an energy shift of both bands
while preserving their relative separation and curvature. In contrast, the latter
governs the nontrivial eigenstate geometry that determines all topological characteristics.

\subsection{Perturbative scenario}\label{IIIA}


As a warm-up, let us consider the perturbative scenario.
Reading from its Hamiltonian~(\ref{eq9}), the vector field $\bm{h}(\bm{k})$ takes the form of
\begin{eqnarray}
\bm{h}(\bm{k})=(2t_x k_x k_y, m, t_z(k_x^2-k_y^2)).\label{Eq_h_pert_model}
\end{eqnarray}
The Berry curvature that is exclusively polarized along the $z$-axis for such two-dimensional system can be
written as~\cite{Saha2016arXiv,Sin2024arXiv,Pires2024arXiv},
\begin{eqnarray}
\Omega_z(\bm{k})=\frac{1}{2|\bm{h}|^3}~\bm{h}\cdot\left(\partial_{k_x}\bm{h}\times\partial_{k_y}\bm{h}\right),
\end{eqnarray}
where $\partial_{k_x}\bm{h} \times \partial_{k_y}\bm{h}$ encodes the local geometric curvature of the $\bm{h}(\bm{k})$
on the momentum-space manifold. Substituting the explicit form of $\bm{h}(\bm{k})$~(\ref{Eq_h_pert_model}),
the Berry curvature simplifies to
\begin{eqnarray}
\Omega_z(\bm{k})=\frac{4t_xt_z m\bm{k}^2}{2|\bm{h}|^3},\label{eq27}
\end{eqnarray}
with $|\bm{h}|=\pm\sqrt{(2t_xk_xk_y)^2+m^2+t_z^2(k_x^2-k_y^2)^2}$.

This implies that the Berry curvature is symmetric under momentum inversion ($\bm{k} \to -\bm{k}$)
and invariant under exchange $k_x \leftrightarrow k_y$.
In addition, its magnitude is proportional to the product of three parameters $t_x t_z m$.
While $m$ modulates the band magnitude, it does not alter the topological properties.
Since the rotational symmetry of the system is governed by the ratio $t_x / t_z $,
this indicates the full rotational symmetry is preserved symmetric at ($t_x = t_z$) as displayed in
Fig.~\ref{bc_typeI_pert}(a), but instead Fig.~\ref{bc_typeI_pert}(b)-(c) show the
rotational symmetry broken for ($t_x \neq t_z$).
In particular, the curvature vanishes at $t_x t_z = 0$, signaling the loss of nontrivial geometry. As long as
$t_x t_z$ is nonzero, the Berry curvature remains finite and undergoes reversal under the sign inversion of $t_x t_z$.

\begin{figure}[htpb]
\centering
\includegraphics[width=2.5in]{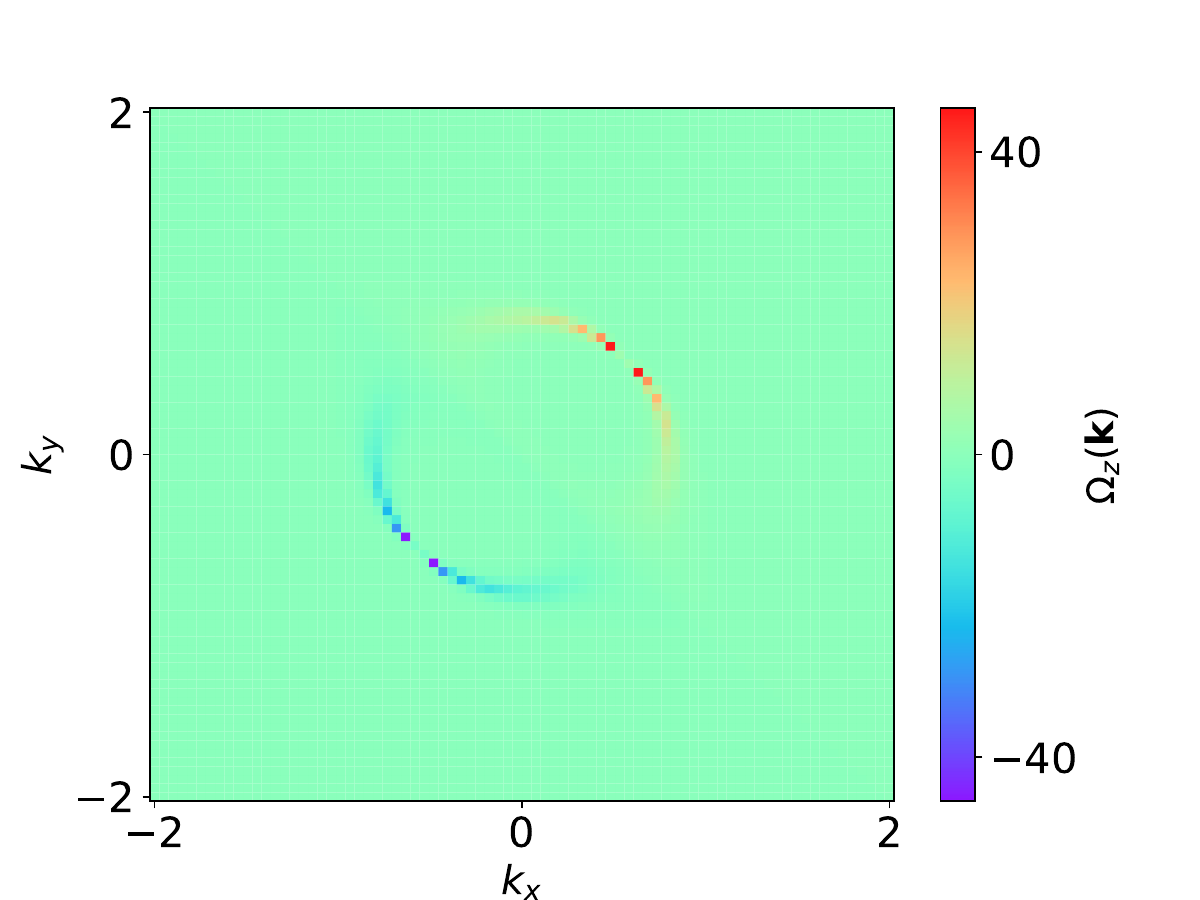}
\vspace{-0.2cm}
\caption{(Color online) Berry curvature of Floquet scenario for the LPL at $\phi=\pi$ and $A_0=0.8,~\hbar\omega=0.43$ with $t_x=t_z=1.4$.}
\label{bc_typeI_Floquet_phi_pi}
\end{figure}

\begin{figure}[htpb]
\centering
\hspace{-0.4cm}
\subfigure[]{\includegraphics[width=1.8in]{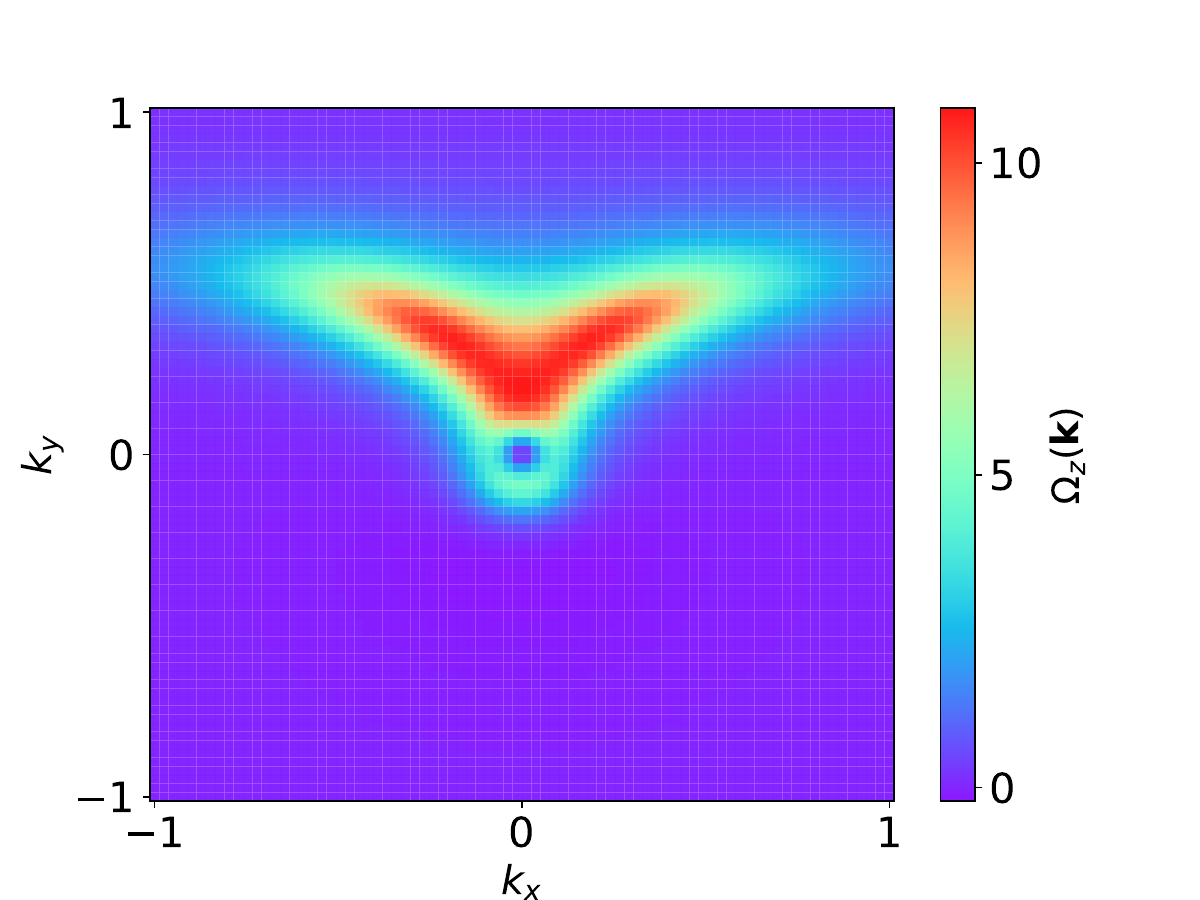}}
\hspace{-0.6cm}
\subfigure[]{\includegraphics[width=1.8in]{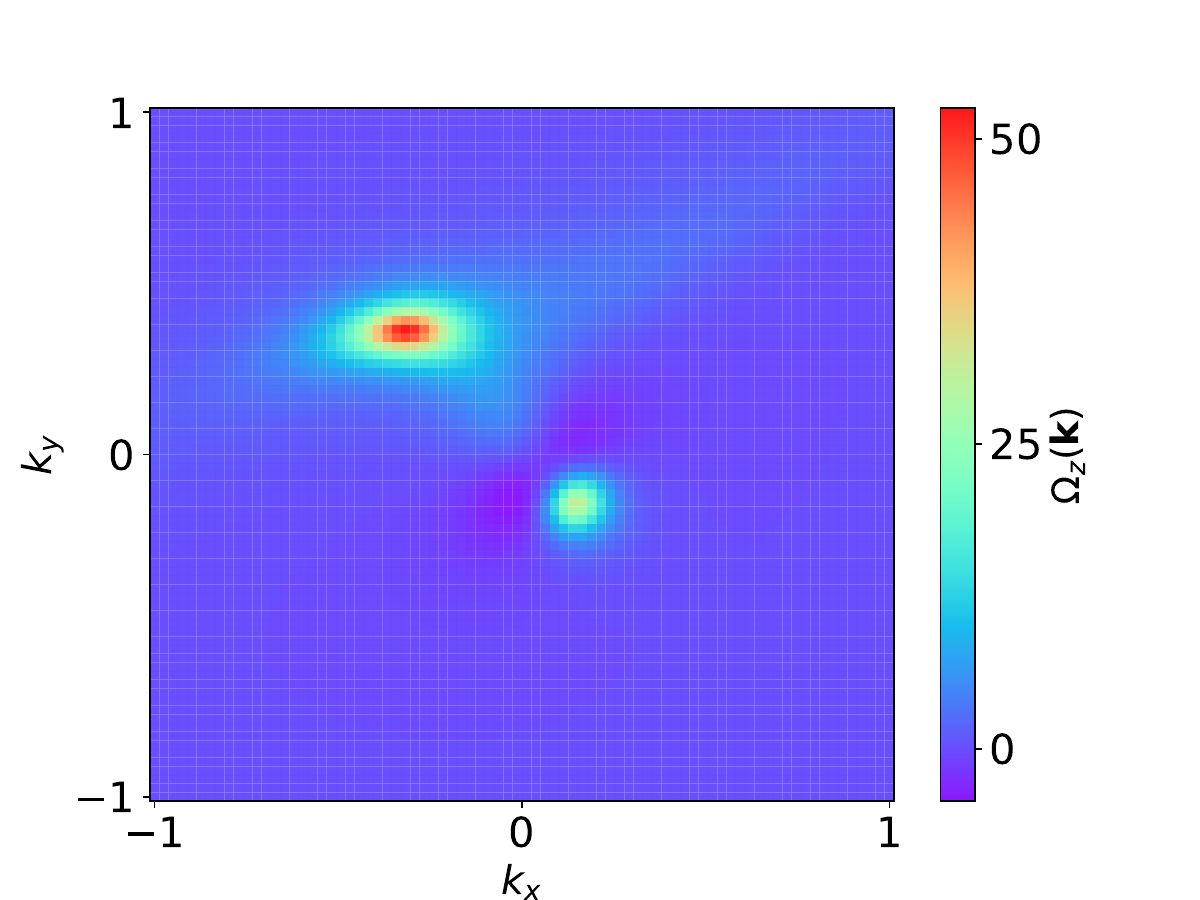}}\\
\hspace{-0.4cm}
\subfigure[]{\includegraphics[width=1.8in]{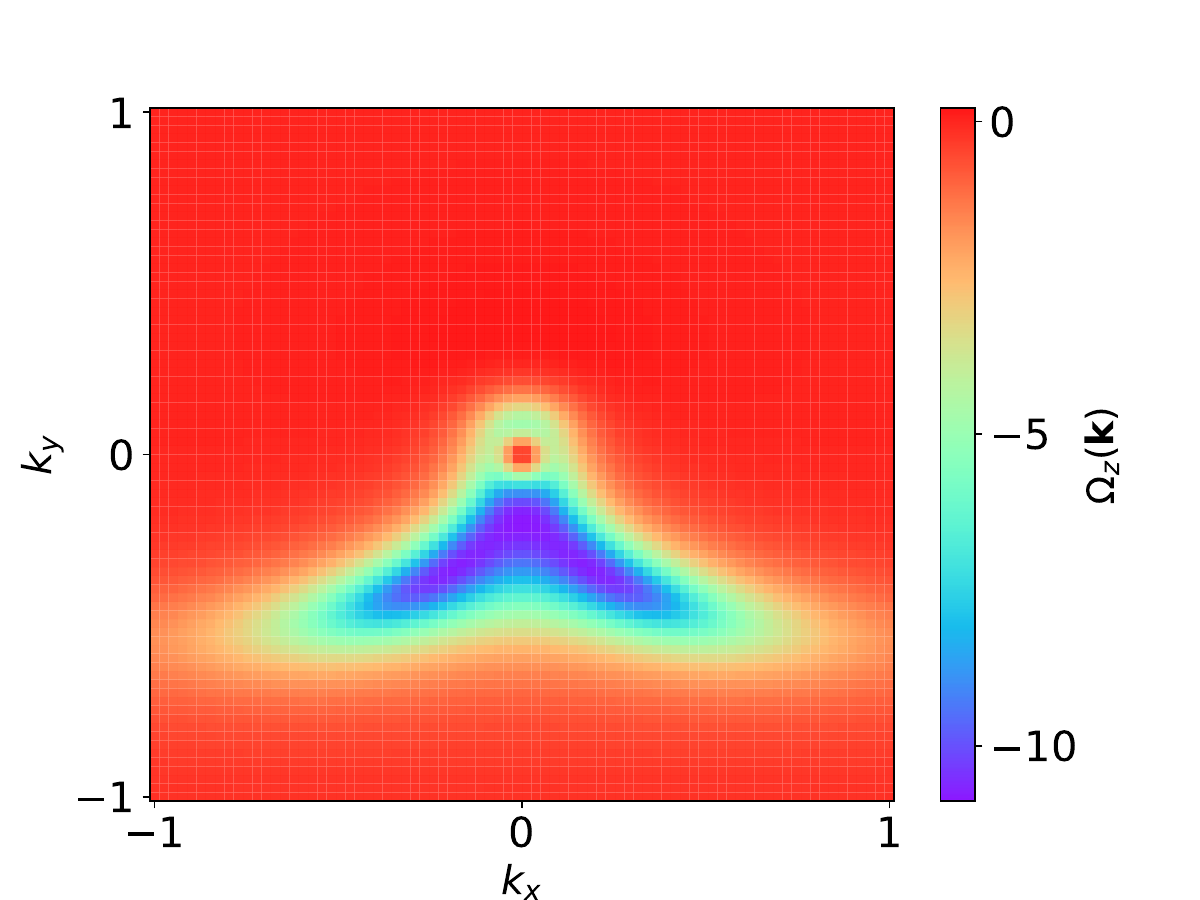}}
\hspace{-0.8cm}
\subfigure[]{\includegraphics[width=1.8in]{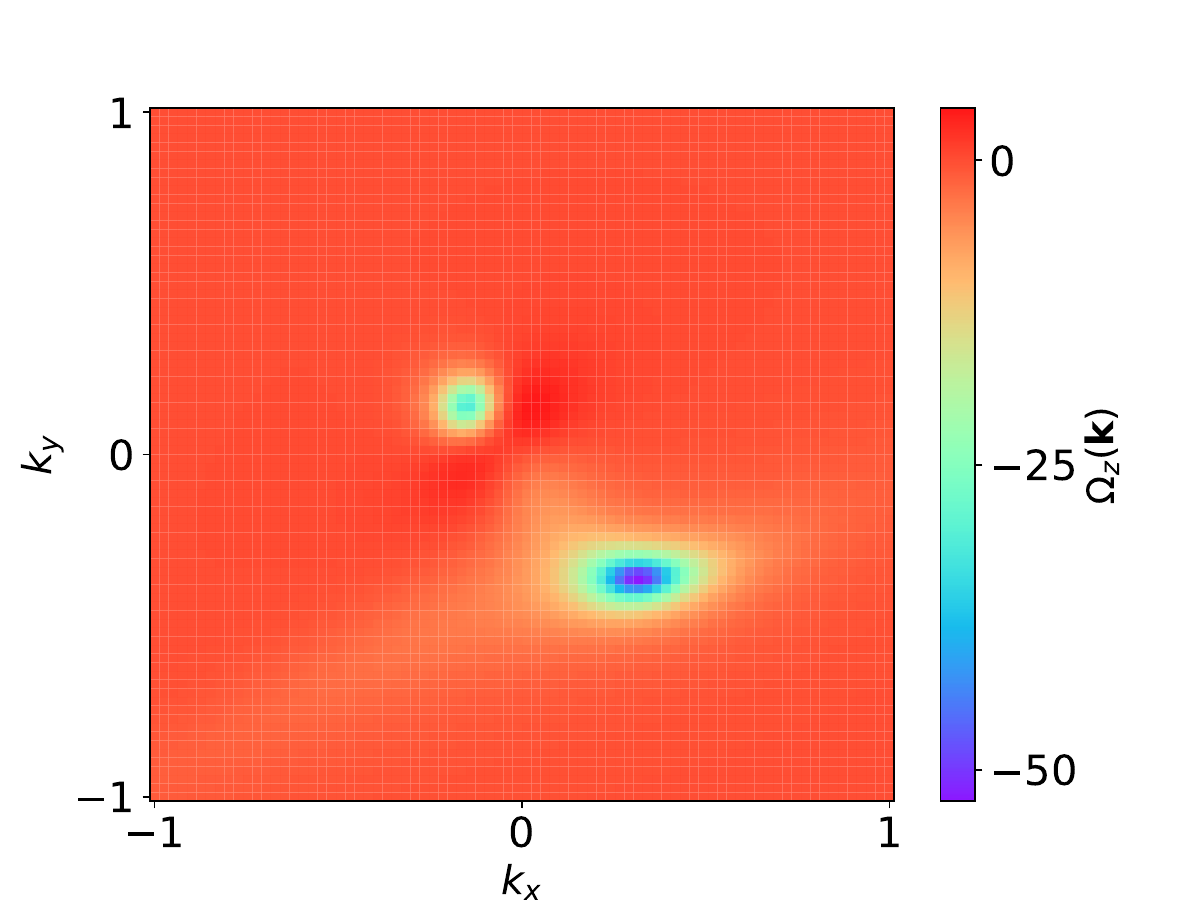}}
\caption{(Color online) Berry curvature of Floquet scenario for the RHPL}: (a) $\phi=\pi/2$ (CPL) and (b) $\phi=\pi/3$ (EPL), as
well as the LHPL: (c) $\phi=-\pi/2$ (CPL) and (d) $\phi=-\pi/3$ (EPL) with $A_0=0.43,~\hbar\omega=0.8$ and $t_x=1.4,~t_z=2.0$.
\label{bc_typeI_Floquet}
\end{figure}

In order to further investigate the potential topological properties,
we need to compute the corresponding Chern number, which is designated by
the integral of the Berry curvature over the Brillouin zone (BZ) \cite{Saha2016arXiv,Xiao2010RMP,Kane2010RMP,Sin2024arXiv,Pires2024arXiv,Singh2024arXiv,Villegas2024arXiv}.
It is then formulated as
\begin{eqnarray}
C=\frac{1}{2\pi}\int_{\mathrm{BZ}}d^2\bm{k}~\frac{4t_xt_z m\bm{k}^2}{2|\bm{h}|^3}.\label{eq28}
\end{eqnarray}
As the Berry curvature of the perturbative scenario~(\ref{eq27})
is invariant under momentum inversion, $\Omega_z(\bm{k}) = \Omega_z(-\bm{k})$,
generally supporting a nontrivial Chern number. Indeed, Eq.~(\ref{eq28}) yields
\begin{equation}
C = \pm \mathrm{sgn}(t_x t_z), \label{eq29}
\end{equation}
where the sign corresponds to the upper and
lower bands shown in Fig.~\ref{Chern}(a), respectively.

This manifestly signifies the perturbative gap transforms a topological transition from the
perturbative QBCP semimetal to a Chern insulator.
Such a quantized result reflects the unique feature of quadratic momentum dependence of the dispersion.
In sharp contrast, the semi-Dirac semimetal with the perturbation term still remains a topological
trivial state ($C=0$)~\cite{Saha2016arXiv,Huang2015PRB}. 
Besides, since the microscopic parameters $t_x$ and $t_z$ illustrated in Fig.~\ref{bc_typeI_pert}
govern both the rotational symmetry of the energy bands and the spatial distribution of the Berry
curvature, the result~(\ref{eq29}) implies that continuous parameter
variations can drive distinct topological phase transitions. A schematic of this dependence
is provided in Fig.~\ref{Chern}(b).

As a consequence, it demonstrates that TRSB via a $\sigma_2$-type perturbation
in a 2D QBCP semimetal can induce a topological phase transition
characterized by a nonzero Chern number.

\vspace{0.3cm}

\subsection{Floquet scenario}\label{IIIB}

Next, let us consider the underlying topological nontrivial properties for the Floquet scenario.
By utilizing the the vector field~(\ref{eq21}) and paralleling the similar derivations of the perturbative scenario,
we arrive at the Berry curvature for the Floquet scenario,
\begin{widetext}
\begin{eqnarray}
\Omega_z(\bm{k})=
\frac{-1}{2|\bm{d}|^3}2t_xt_z\left[\bm{k}^2(k_x\partial_{k_x}m_{\mathrm{eff}}
+k_y\partial_{k_y}m_{\mathrm{eff}}-2m_{\mathrm{eff}})+e^2A_0^2(k_y\partial_{k_x}m_{\mathrm{eff}}
+k_x\partial_{k_y}m_{\mathrm{eff}})\cos\phi\right].\label{eq30}
\end{eqnarray}
\end{widetext}
where $\bm{d}$ is designated in Eq.~(\ref{eq21}). It can be observed that the Berry curvature $\Omega(\bm{k})$ explicitly depends on both the microscopic parameters $t_{x,z}$ and the characteristics of the applied electric field including its amplitude $A_0$, frequency $\hbar\omega$, and polarization angle $\phi$.

Reading off Eq.~(\ref{eq30}), it can be found that the polarization angle $\phi$ is crucial
to modulate the Berry curvature by controlling the multitude of the second term, which scales as $\cos\phi$.
To systematically address this issue, let us consider all three distinct polarization states
mentioned in Sec.~\ref{Subsubsec_Floquet}.

With respect to the LPL at $\phi = 0, \pi$ displayed in Fig.~\ref{bc_typeI_Floquet_phi_pi},
it considerably suppresses the Berry curvature but preserves the central symmetry about $\bm{k}=0$, namely
$\Omega_z(k_x, k_y) = -\Omega_z(-k_x, -k_y)$.
As for the CPL and EPL, we firstly consider the right-handed polarized light (RHPL)
in which the $\phi$ is restricted to $(0,\pi)$.
Considering the CPL with $\phi = \pi/2$, the second term in Eq.~(\ref{eq30}) vanishes. This indicates that the
Berry curvature possesses the mirror symmetry about the $k_x$-axis, i.e., $\Omega_z(k_x)=\Omega_z(-k_x)$,
while breaking rotational and $k_y$-mirror symmetries as depicted in Fig.~\ref{bc_typeI_Floquet}(a).
Turning to the EPL, Fig.~\ref{bc_typeI_Floquet}(b) presents the behavior of $\Omega_z(k_x,k_y)$ at
as a representative value of $\phi=\pi/3$ (The basic result is similar for other values.).
Compared to its CPL counterpart, one can notice from Fig.~\ref{bc_typeI_Floquet}(b)
that the EPL induces complete symmetry breaking in momentum space:
$\Omega_z(k_x, k_y) \neq \Omega_z(-k_x, k_y) \neq \Omega_z(k_x, -k_y)$.
In addition, we have also examined the left-handed polarized light (LHPL) with $\phi$ located at $(-\pi,0)$. Fig.~\ref{bc_typeI_Floquet}(c)-(d) suggest that the LHPL shares the analogous results with RHPL but only
satisfies $\Omega_z^{\mathrm{LHPL}}(\bm{k}) = -\Omega_z^{\mathrm{RHPL}}(\bm{k})$ for both CPL and EPL.

\begin{figure}[htpb]
\centering
\includegraphics[width=3in]{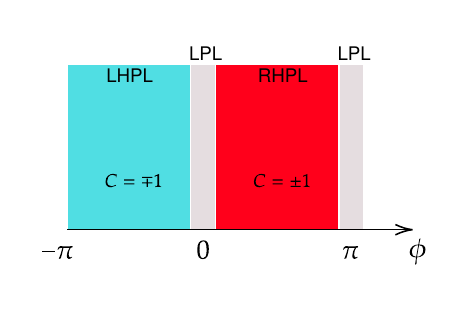}
\vspace{-1cm}
\caption{(Color online) The polarization position $\phi$-dependent Chern numbers for the Floquet scenario. Hereby,
the LPL corresponds to topologically trivial state with $C=0$.}
\label{phasediagram2}
\end{figure}

In addition to the polarization angle, we find that the qualitative behavior of Berry curvature exhibits weak dependence on
other parameters in Eq.~(\ref{eq30}) such as the light intensity $A_0$, its frequency $\hbar\omega$, and the microscopic structural parameters $t_x,~t_z$. The essential reason is that these parameters can only primarily modulate the magnitude
of energy gap but are not able to alter either the topological invariants or symmetry constraints. Particularly,
the Berry curvature of the Floquet scenario~(\ref{eq30}) scales as $\Omega_z\propto(t_x t_z)^2$
rendering it invariant under sign reversals of $t_x$ or $t_z$, which is qualitatively
different from its counterpart of the perturbative scenario~(\ref{eq27}).

To proceed, we are able to discuss the Chern number. As the qualitative behavior of Berry curvature is insensitive to the
parameters $A_0$, $\hbar\omega$, and $t_x, t_z$  in Eq.~(\ref{eq30}), they do not alter the topological invariants and cannot influence
the topological phase transition. However, the LPL at $\phi = 0, \pi$ causes $\Omega_z(\bm{k})=-\Omega_z(-\bm{k})$ and thus
yields a trivial Chern number $C=0$. This signals that the LPL opens an energy gap in the 2D QBCP system, leaving the system in a conventional insulating phase.

In sharp contrast, the application of nonlinear polarization (non-LPL, i.e., CPL or EPL) with $\phi \neq 0, \pi$
fundamentally alters the symmetry of Berry curvature, resulting in $\Omega_z(\bm{k})\neq-\Omega_z(-\bm{k})$.
Paralleling the strategy in Sec.~\ref{IIIA}, we find that
non-LPL irradiation not only removes the QBCP by opening a energy gap but also induce a topological insulator state,
which owns nontrivial Chern numbers $C=\pm 1$ with the sign denoting the upper and lower bands, respectively.

Specifically, the polarization direction (PD) of the light directly governs the sign of the Berry curvature and
henceforth determines the Chern number inversion between distinct energy bands:
RHPL with $0 < \phi < \pi$ (CPL or EPL) causing a positive Berry curvature $\Omega_z > 0$ and leading to $C = 1$ to the upper band,
and its LHPL counterpart ($-\pi < \phi < 0$) giving rise to a negative Berry curvature and
the Chern number $C = -1$ for the upper band. This implies that the system undergoes
$C_{\text{upper}} \xrightarrow{\phi: \, 0 \to -\pi} 1 \to -1,  \quad C_{\text{lower}} \xrightarrow{\phi: \, 0 \to -\pi} -1 \to 1$
with tuning the parameter $\phi$ from RHPL to LHPL. In other words, the sign of PD $\mathrm{sgn}(\phi)$
acts as a tunable order parameter to control a topological phase transition between different topological
invariants as schematically shown in Fig.~\ref{phasediagram2}.

To wrap up, an irradiation of a 2D QBCP semimetal with a polarization tunable light can open an energy gap and break time-reversal symmetry (TRS), thereby inducing nontrivial and topological properties as summarized in Table.~\ref{cn_tab}. In particular,
a potential topological phase transition can be induced between a Chern insulator and trivial insulator
with variation of the PD as displayed in Fig.~\ref{phasediagram2}.

\begin{table}[htpb]
\centering
\caption{Collections of Chern number for the Floquet scenario under distinct polarization directions (PDs).}
\vspace{0.3cm}
\begin{tabular}{c|c c c c} 
\hline
\hline
 Topological Invariant & LPL & CPL & EPL &  PDs\\
\hline
Chern number & 0 & $\pm 1$ & $\pm 1$ & RHPL\\
\hline
Chern number & 0 & $\mp 1$ & $\mp 1$ & LHPL\\
\hline
\hline
\end{tabular}
\label{cn_tab}
\end{table}


\subsection{Hybrid situation}\label{IIIC}

For completeness, let us study the hybrid situation in which it is irradiated by a polarization
light imposed in the Floquet scenario~(\ref{eq16}) on the perturbative scenario~(\ref{eq9}).
In this circumstance, the Berry curvature is reformulated as
\begin{widetext}
\begin{eqnarray}
\Omega_z(\bm{k})
=
\frac{-1}{2|\bm{d}|^3}2t_xt_z\left[\bm{k}^2(k_x\partial_{k_x}m_{\mathrm{eff}}
+k_y\partial_{k_y}m_{\mathrm{eff}}-2m_{\mathrm{eff}}-2m)+e^2A_0^2(k_y\partial_{k_x}m_{\mathrm{eff}}
+k_x\partial_{k_y}m_{\mathrm{eff}})\cos\phi\right],\label{eq37}
\end{eqnarray}
where the magnitude of vector field is designated as
\begin{eqnarray}
|\bm{d}(\bm{k})|=\pm\sqrt{(t_xe^2A_0^2\cos\phi+2t_xk_xk_y)^2+(m_{\mathrm{eff}}+m)^2+t_z^2(k_x^2-k_y^2)^2}.
\end{eqnarray}
\end{widetext}

\begin{figure}[htpb]
\centering
\includegraphics[width=3.5in]{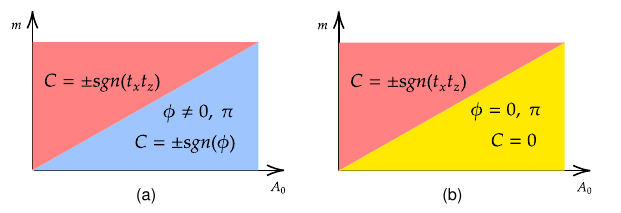}
\vspace{-0.8cm}
\caption{(Color online) Schematic $m-A_0$ dependent phase diagram of the Hybrid situation for
(a) non-LPL (CPL or EPL) with $\phi\neq 0,~\pi$ and (b) LPL at $\phi=0,~\pi$.}
\label{phasediagram3}
\end{figure}

\begin{figure*}[htpb]
\centering
\subfigure[]{\includegraphics[width=2.2in]{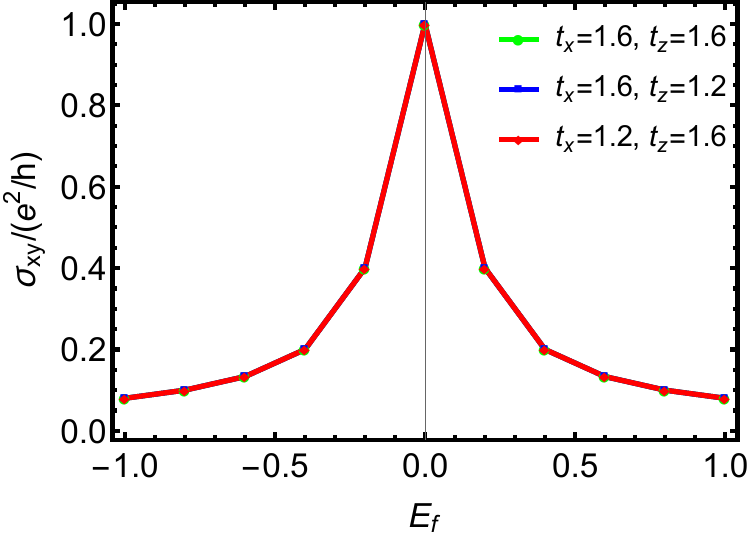}}
\subfigure[]{\includegraphics[width=2.2in]{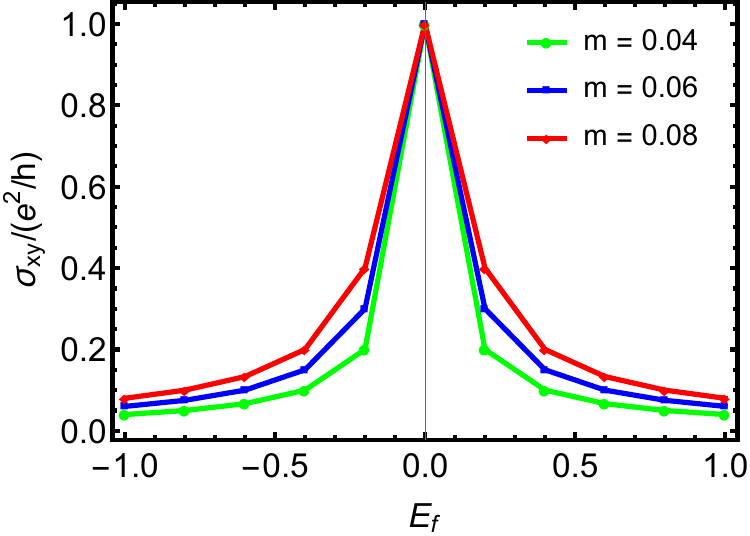}}
\subfigure[]{\includegraphics[width=2.2in]{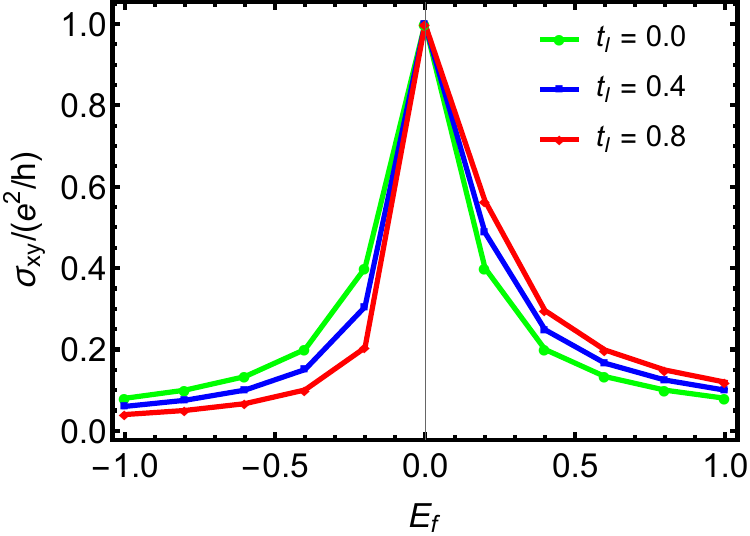}}\\
\vspace{-0.3cm}
\caption{(Color online) The $E_f$-dependent AHC $\sigma_{xy}/(e^2/\hbar)$ of the perturbative scenario for
(a) $m=0.08$ and $t_I=0.0$ with variations of $t_x,~t_z$,  (b) $t_x=t_z=1.6,~m=0.08$ with $t_I\neq0$,
and (c) $t_x=t_z=1.6,~t_I=0.0$ with distinct values of $m$.}
\label{AHC_pert}
\end{figure*}

Either the static symmetry-breaking perturbation term or the dynamic Floquet polarization
light is able to open an energy gap and break TRS. However, the former is primarily dependent upon the
parameter $m$~(\ref{Eq_h_pert_model}) but instead the latter dictates the topological symmetry properties
via polarization angle $\phi$~(\ref{eq30}). It is therefore motivates us to study their concomitant effects
on driving topological phase transitions.

In the regime where the intensity of polarization light dominates over the perturbation
term with $A_0 \gg m$, the Berry curvature is qualitatively consistent with those of Floquet mode presented
in Figs.~\ref{bc_typeI_Floquet_phi_pi}-\ref{bc_typeI_Floquet} and thus not shown hereby. Accordingly, the topological
properties are again dictated by the polarization angle $\phi$ and the Chern numbers can be
summarized as follows
\begin{eqnarray}
C=
\begin{cases}
\pm\mathrm{sgn}(\phi),~\phi\neq 0,\pi\\
0,~\phi=0,\pi,
\end{cases}
\end{eqnarray}
which manifestly reflect the polarization-controlled topology.
In comparison, the static perturbation term governs the behavior of Berry curvature
when the light intensity is comparable to or smaller than
the perturbative mass ($A_0 \leq m$). Under this circumstance, the structural parameters $t_x$ and $t_z$ replace
the polarization angle $\phi$ to determine the Chern number as $C = \pm\mathrm{sgn}(t_x t_z)$.
This indicates the topological phase transition is dictated solely by the microscopic structural parameters
$t_x$ and $t_z$. Fig.~\ref{phasediagram3} schematically summarizes the central conclusions for the
Hybrid situation.


\section{Anomalous Hall conductivity}\label{Sec_Hall_conductivity}

Subsequently, we are going to examine the behavior of quantum transport accompanied by the topological phase transition.
The anomalous Hall conductivity (AHC)~\cite{Karplus1954PR,Streda2010PRB,Saha2016arXiv,Qi2006PRB,Nandy2024arXiv,Habe2025PRB,Baidya2025arXiv} represents a fundamental signature of momentum-space topology, stemming from the geometric
Berry curvature $\Omega_z(\bm{k})$ intrinsic to the electronic band structure.
Distinct from conventional Hall effects~\cite{Hall1879AJOM}, which require external magnetic
fields (ordinary Hall effect) or Landau-level quantization (quantum Hall effect)~\cite{Klitzing1986RMP},
the AHC originates solely from the momentum-space Berry phase accumulation in crystalline materials~\cite{Thouless1982PRL,Haldane1988PRL}.
It is therefore expected to emerge a finite transverse conductivity $\sigma_{xy}$ in our models at zero external field,
serving as an experimental signature of symmetry-protected topological phases.

The transverse AHC is formally expressed as~\cite{Saha2016arXiv,Qi2006PRB,Nandy2024arXiv,Habe2025PRB,Baidya2025arXiv}
\begin{eqnarray}
\sigma_{xy}=\frac{e^2}{\hbar}\sum_n\int_{\mathrm{BZ}}\frac{d^2\bm{k}}{(2\pi)^2}f(E^n(\bm{k}))\Omega_z^n(\bm{k}),\label{eq40}
\end{eqnarray}
where the summation $n$ spans all energy bands, the integration covers the Brillouin zone (BZ), and $\Omega_z^n(\bm{k})$ denotes the Berry curvature of the $n$-th band. Hereby, the Fermi-Dirac distribution is given by
\begin{eqnarray}
f(E)=\frac{1}{1+e^{(E-E_f)/k_BT}},
\end{eqnarray}
with $E_f$ defining the Fermi energy. At the zero-temperature limit ($T \to 0$), the $f(E)$ reduces to a step function,
\begin{eqnarray}
f(E(\bm{k}))\to\Theta(E(\bm{k})-E_f),
\end{eqnarray}
effectively restricting the integration to states below $E_f$.
Generally, the AHC becomes quantized at temperatures where thermal excitations are negligible
($k_B T \ll \Delta_{\mathrm{gap}}$). In the following, let us systematically analyze the behavior of AHC for the scenarios
introduced in Sec.~\ref{Sec_model}.


We commence with studying the perturbative scenario in Sec.~\ref{subsubSec_toy_model}.
Following Eq.~(\ref{eq40}) and assuming the equilibrium-state $f(E)$, the numerical results in Fig.~\ref{AHC_pert}
present the Fermi level $E_f$ dependence of the AHC.
It clearly shows that the AHC saturates to the quantized value at $E_f=0$,
$\sigma_{xy} = \frac{Ce^2}{\hbar}, \quad C = 1$, which is well consistent with
the Thouless-Kohmoto-Nightingale-den Nijs (TKNN) formula~\cite{Thouless1982PRL}.
For $E_f\neq 0$, $\sigma_{xy}$ may be also sensitive to other parameters.
As the parameter determines the PH symmetry and $m$ adjusts the energy gap of the system, they provide
a quantitative impact on the AHC as depicted in Fig.~\ref{AHC_pert}(a) and Fig.~\ref{AHC_pert}(b), respectively.
In particular, $t_I\neq 0$ can break particle-hole (PH) symmetry, inducing an asymmetric AHC on distinct sides of $E_f=0$
and the AHC is proportional to $m$. However, the parameters $t_x,~t_z$ shown in Fig.~\ref{AHC_pert}(c) only
energy band symmetry but do not alter the magnitude of AHC.

\begin{figure}[htpb]
\centering
\subfigure[]{\includegraphics[width=1.6in]{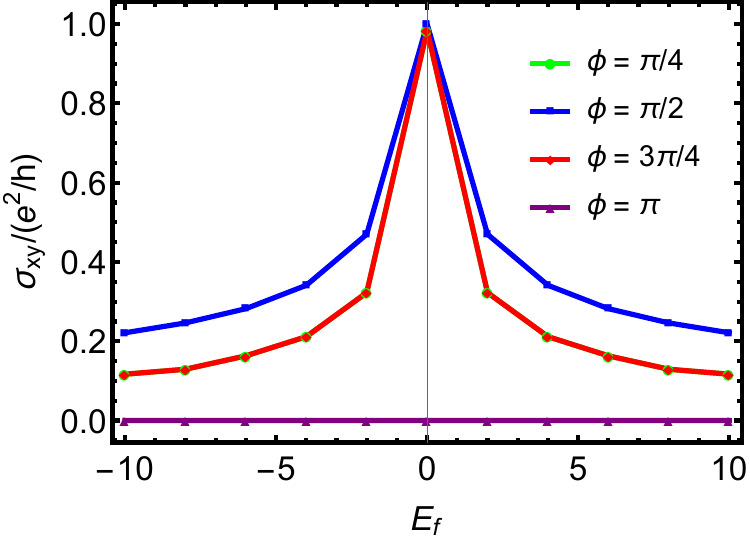}}
\subfigure[]{\includegraphics[width=1.6in]{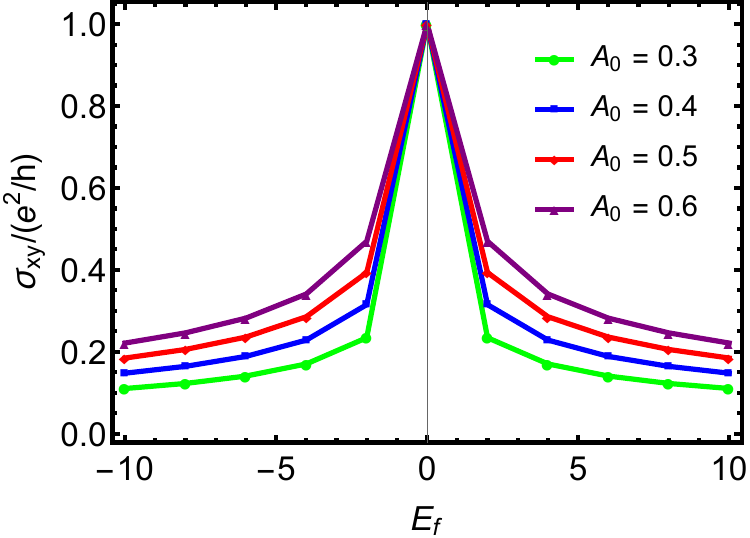}}
\subfigure[]{\includegraphics[width=1.6in]{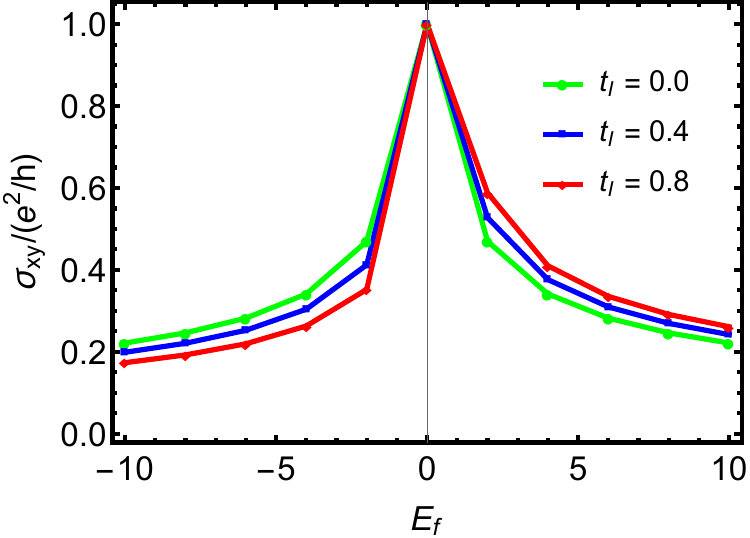}}
\subfigure[]{\includegraphics[width=1.6in]{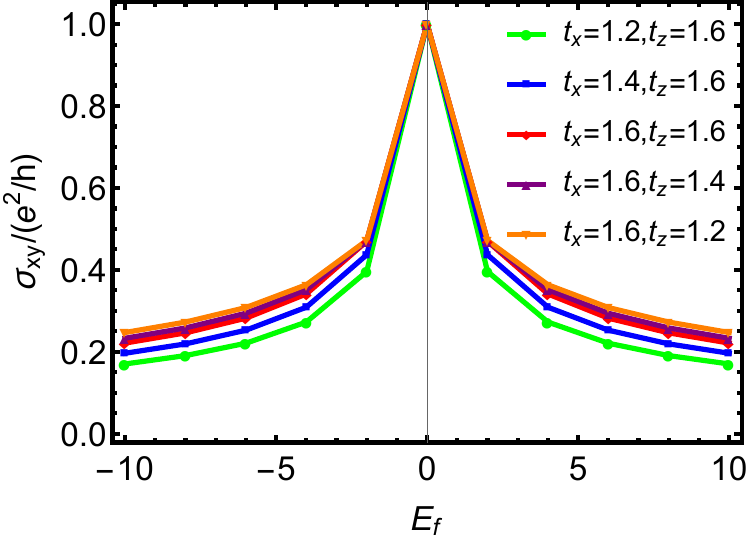}}\\
\vspace{-0.2cm}
\caption{(Color online) The $E_f$-dependent AHC $\sigma_{xy}/(e^2/\hbar)$ of the perturbative scenario for
(a) distinct values of $\phi$ with $t_x=t_z=1.6,~A_0=0.6,~\hbar\omega=0.43$ at $t_I=0.0$,
(b) distinct values of $A_0$ with $t_x=t_z=1.6,~\phi=\pi/2,~\hbar\omega=0.43$ at $t_I=0.0$,
(c) $\phi=\pi/2,~t_x=t_z=1.6,~A_0=0.6,~\hbar\omega=0.43$ with variations of a finite $t_I$,
and (d) distinct values of  $t_x,~t_z$ with $\phi=\pi/2,~A_0=0.6,~\hbar\omega=0.43$ at $t_I=0.0$.}
\label{AHC_Floq}
\end{figure}

Next, we move to consider the AHC of the Floquet scenario~\ref{Subsubsec_Floquet},
which is characterized by the polarization angle and strength of external light as well as microscopic structural parameters.
As aforementioned in Sec.~\ref{IIIB}, the nontrivial the Berry curvature and the Chern number are
primarily determined by the polarization angle of external light.
We therefore begin with investigating the influence of $\phi$ on the AHC as presented in Fig.~\ref{AHC_Floq}.
Reading from Fig.~\ref{AHC_Floq}(a), the LPL with $\phi=0,~\pi$ generates negligible Berry curvature ($\Omega_z(\bm{k}) \approx 0$)
across the Brillouin zone, yielding $\sigma_{xy} \to 0$ and preserving the topologically trivial insulating phase.
In comparison, for either CPL ($\phi=\pm\pi/2$) or EPL ($\phi \neq 0, \pi,\pm\pi/2$) can
induce nonzero Berry curvature with TRSB and drive a quantized AHC $\sigma_{xy} = e^2/\hbar$ that
is consistent with the behavior of Chern insulator
when approaching the Fermi energy $E_f$. This indicates that the polarization-dependent symmetry
breaking plays an essential role in switching trivial and topological insulating phases.

Given the qualitative similarity between CPL and EPL responses, let us subsequently take the CPL as an example to investigate
the effects of other parameters on AHC. The strength of external field $A_0$ significantly modifies the energy
gap despite it cannot qualitatively alter the Berry curvature.
This establishes a positive correlation as shown in Fig.~\ref{AHC_Floq}(b),
where increasing $A_0$ enhances the AHC near the Fermi energy.
Learning from Fig.~\ref{AHC_Floq}(c), one can find that a finite $t_I$ breaks PH symmetry and
induces an asymmetric AHC about $E_f = 0$ with the asymmetry amplitude proportional to
$|t_I|$. In contrast, $t_x$ and $t_z$ exhibit more complex influences as depicted in Fig.~\ref{AHC_Floq}(d). They not only govern rotational symmetry but also quantitatively determine both the magnitude of energy gap,
which jointly quantitatively modifying the AHC.

Furthermore, let us briefly comment on the AHC of the hybrid situation.
In such case, the perturbation parameter $m$ competes with the Floquet driving field $A_0$.
When $A_0$ is dominant at $A_0 \gg m$, the AHC exhibits analogous behavior to those shown
in Fig.~\ref{AHC_Floq}. Conversely, while $m$ dominates over light-induced effects, the AHC behavior is restored to those of the perturbative scenario presented in Fig.~\ref{AHC_pert}.

To recapitulate, the AHC acts as a hallmark of topological states,
arising from TRSB and nontrivial Berry curvature,
and serves as a key experimental observable in the linear-response regime,
offering a direct and quantifiable signature of the nontrivial Chern number.
We analyze two distinct gap-opening mechanisms, both of which break TRSB and thus generate nontrivial Chern number,
yielding the quantized AHC. The key conclusions are provided in Fig.~\ref{AHC_pert} and Fig.~\ref{AHC_Floq},
which reveal universal scalings of magnitude and symmetry for the AHC near $E_f$ with variations of interaction parameters.

\section{Optical signatures of topological phase transitions}\label{Sec_optical}

The preceding sections detail the TRS-breaking transition from a QBCP semimetal to a Chern insulator
via static and Floquet scenarios, identifying the quantized AHC as an experimental signature. Motivated
by recent studies indicating that the free QBCP semimetal possesses a non-Abelian Euler invariant $\chi=1$~\cite{Jankowski2025PRB,Jain2025PRB}, we hereby complement our work with a brief Euler-class
discussion and assess the potential of optical methods to probe these transitions.

\subsection{Euler-class invariant of the free QBCP system}\label{Sec_Euler-class}

Let us focus on the free QBCP system described by Eq.~(\ref{H_01}). Its protected band degeneracy
stems from the underlying either $\mathcal{PT}$ or $\mathcal{C}_2\mathcal{T}$ symmetry.
These symmetries are directly responsible for the emergence of a non-Abelian topological state
characterized by an Euler class~\cite{Jankowski2025PRB}, which yields a topological invariant
that differs essentially from the Abelian Chern number.

To derive this invariant, we follow the approach in Ref.~\cite{Bouhon2020NP}. The Euler connection
and the Euler form are denominated by
\begin{eqnarray}
\bm{a}(\bm{k}) \equiv \langle u^1(\bm{k}) | \nabla_{\bm{k}} u^2(\bm{k}) \rangle,\,\,\,
\mathrm{Eu}(\bm{k}) \equiv \nabla_{\bm{k}} \times \bm{a}(\bm{k}),
\end{eqnarray}
where the real eigenstates for the lower and upper bands can be chosen as,
\begin{eqnarray}
|u^1(\mathbf{k})\rangle \!=\!\! \left(\!\begin{array}{c} -\sin(\gamma/2) \\ \cos(\gamma/2) \end{array}\!\right),
|u^2(\mathbf{k})\rangle \!=\!\! \left(\!\begin{array}{c} \cos(\gamma/2) \\ \sin(\gamma/2) \end{array}\!\right),
\end{eqnarray}
with
\begin{eqnarray}
\cos\gamma \equiv \frac{d_z}{|\bm{d}|}, \quad \sin\gamma \equiv \frac{d_x}{|\bm{d}|},
\quad \tan\gamma \equiv \frac{t_x}{t_z} \tan(2\theta).
\end{eqnarray}
Then we integrate $\bm{a}$ along the boundary $\partial D$
of a disk $D$ centered at the origin point. After several calculations, the Euler class on the region $D$
can be expressed as~\cite{Jankowski2025PRB,Bouhon2020NP,Chau2025PRB},
\begin{eqnarray}
\chi(D)
&=& \frac{1}{2\pi} \left[ \int_D \mathrm{Eu}(\bm{k}) \, d^2\bm{k} - \oint_{\partial D}
\bm{a}(\bm{k}) \cdot d\bm{k} \right]\nonumber\\
&=& -\mathrm{sgn}(t_x t_z),
\end{eqnarray}
which leads to $\chi = 1$ with ignoring the overall sign convention~\cite{Jankowski2025PRB}.
This confirms that the free Hamiltonian Eq.~(\ref{H_01}) carries a nontrivial patch Euler class $\chi=1$, which is in
qualitative agreement with the characterization of quadratic Euler nodes in Ref.~\cite{Jankowski2025PRB}.

In this sense, the QBCP semimetal can be identified as a non-Abelian Euler semimetal.
Breaking its $\mathcal{C}_2\mathcal{T}$ symmetry through either static perturbation or dynamic Floquet driving
induces a topological phase transition to a Chern insulator. This process fundamentally converts nodal bands
with Euler invariant \(|\chi| = 1\) into gapped bands carrying opposite Chern numbers
$C=\pm 1$~\cite{Jankowski2025PRB,Bouhon2022arXiv}. Thus, it demonstrates the controlled conversion
of a non-Abelian Euler semimetal into an Abelian Chern insulator through the symmetry breaking.

\subsection{Optical deduction of topological phases}

The distinct topology of each phase gives rise to a characteristic optical
signature~\cite{Jankowski2025PRB, Chau2025PRB}. Specifically, the QBCP semimetal
with $\chi=1$ shows non-chiral features, while the Chern insulator exhibits strong
circular dichroism (CD) due to its broken TRS and chiral electronic band
structure~\cite{Jankowski2025PRB, Chau2025PRB}. This sharp contrast makes CD
spectroscopy serves as a sensitive experimental tool for identifying and
tracking the topological phase transition, which complements the quantized
AHC discussed in Sec.~\ref{Sec_Hall_conductivity}.

\begin{table*}[htbp]
  \centering
  \caption{Optical signatures for distinguishing topological phase transitions in QBCP materials (CI: Chern insulator, NLP/LP:
nonlinear/linear polarization, $\Delta$: band gap, SM: semimetal, and TTI: topological trivial insulator)~\cite{Jankowski2025PRB, Chau2025PRB,Tran2017SA,Asteria2019NP}.}\label{tab_II}
  \vspace{0.3cm}
  \begin{tabular} {l|c c c c c}
  \hline
  \hline
  \hspace{0.3cm}\textbf{Phase} & \textbf{TRS} & \textbf{Topological Invariant} & \textbf{CD} & \textbf{Optical Weight} & \textbf{Higher-Order Photoconductivity} \\
  \hline
  QBCP SM & Preserved & Euler class $\chi=1$ & Negligible ($\sim 0$) & $W \geq |\chi|e^2/h$ & Significant, ratio related to $\chi$ \\
  \hline
   CI (static) & Broken & $C=\pm\text{sgn}(t_x t_z)$ & Strong & Gap-constrained ($\propto \Delta$) & May be present, non-dominant feature \\
  \hline
   CI (NLP) & Broken & $C=\pm\text{sgn}(\phi)$ & Strong & Parameter-dependent & Typically suppressed by driving \\
  \hline
  TTI (LP) & Broken & $C=0$ & Negligible ($\sim 0$) & Parameter-dependent & May be observable \\
  \hline
  \hline
  \end{tabular}
  \end{table*}

To elucidate this issue, we compare the distinct features of CD between the QBCP semimetal and
the Chern insulator. As to the QBCP semimetal, it preserves TRS and its Hamiltonian
is protected by $\mathcal{C}_2\mathcal{T}$ or $\mathcal{PT}$ symmetry. This results
in a vanishing Berry curvature and a negligible CD response.
However, its nontrivial Euler invariant $\chi$ exhibits
distinct optical property~\cite{Jankowski2025PRB,Chau2025PRB},
which is dominated by the quantum metric with the optical weight bounded by
$W_{aa}(\omega_{\mathrm{max}}) \ge |\chi| e^2/h$. While lacking a linear chiral response,
the system exhibits higher-order nonlinear effects. For instance, the third-order jerk photoconductivity reads $\sigma_\mathrm{jerk}^{xxxx}/\sigma_\mathrm{jerk}^{xxyy}=(3|\chi|-1)/(|\chi|+1)$~\cite{Jankowski2025PRB,Chau2025PRB},
which is tied to $\chi$ and independent of optical chirality.

With respect to the Chern insulator, it is characterized by a nonzero Chern number and an
asymmetric Berry curvature, which together produce a distinct chiral optical response.
For the static scenario, the difference in absorption for left- and right-circularly polarized light
$\Delta\alpha(\omega) = \alpha_+(\omega) - \alpha_-(\omega) \neq 0$ is nonzero~\cite{Tran2017SA,Asteria2019NP}.
In the low-frequency limit, the CD quantizes as $\Delta\alpha \propto A^2 C / \hbar$,
directly linking its sign to the Chern number $C$~\cite{Tran2017SA,Asteria2019NP}.
Under Floquet driving, the polarization of the light serves a dual role as both a control
parameter and a probe. As illustrated in Fig.~\ref{phasediagram2}, the circular polarization
($\phi = \pm \pi/2$) drives the system into a Chern insulator with $C = \pm 1$
and generates a strong CD signal, while linear polarization ($\phi = 0, \pi$) results
in a trivial insulator with vanishing CD. In particular, measuring the circular dichroism spectrum thus
provides a direct experimental strategy to identify the Chern insulator phase and determine the sign
of its Chern number.

As a corollary, the distinct optical signatures of each topological phase offer a feasible
guidance for experimental observations. The key optical observables for each phase summarized
in Table~\ref{tab_II}, providing clear optical properties
for identifying the QBCP semimetal, the Chern insulator, and the trivial insulating phase.

\section{Summary}\label{Sec_summary}

In summary, we conduct a systematic investigation of nontrivial topological states emerging
in 2D QBCP semimetals by introducing two complementary scenarios: the static perturbative and the dynamic
Floquet scenarios, both designed to break TRS~\cite{Saha2016arXiv,Chong2008PRB,
Polkovnikov2014arXiv,Benhaida2024arXiv,Kibis2024arXiv,Asgari2025arXiv,
Oka2009PRB,Rudner2020NRP,Liu2025arXiv,Fu2025arXiv,Yokoyama2025arXiv}.
The detailed analysis reveals that they produce different
distributions of Berry curvature in momentum space. These accordingly give rise to nontrivial
topological states and then induce distinct topological phase transitions via tuning their related parameters.

As for the static perturbative scenario, it eliminates the QBCP and opens an energy gap with the TRSB.
The Berry curvature displayed in Fig.~\ref{bc_typeI_pert} exhibits symmetric under momentum inversion ($\bm{k}\to-\bm{k}$),
leading to the emergence of Chern insulator characterized by Chern number $C = \pm \mathrm{sgn}(t_x t_z)$
that is dependent on structural parameters $t_x$ and $t_z$. Fig.~\ref{Chern} schematically presents how
continuous variation of these coupling parameters drives discrete transitions between topologically distinct states.

Considering the Floquet scenario, the optical polarization angle ($\phi$) plays an essential role
in governing both the symmetry properties and topological character of Berry curvature as demonstrated
in Fig.~\ref{bc_typeI_Floquet}. As depicted in Fig.~\ref{bc_typeI_Floquet_phi_pi}, the LPL at $\phi=0$ or $\pi$
preserves the central antisymmetry of the Berry curvature, resulting in a topologically trivial phase with Chern number.
Conversely, either CPL or EPL breaks this antisymmetry of Berry curvature,
yielding $\Omega_z(\bm{k})\neq-\Omega_z(-\bm{k})$ shown in Fig.~\ref{bc_typeI_Floquet}.
This results in a nontrivial Chern insulator with Chern numbers $C = \pm\mathrm{sgn}(\phi)$,
where $\phi>0$ and $\phi<0$ correspond to RHPL and LHPL, respectively. This indicates that $\phi$ displayed in
Fig.~\ref{phasediagram2} acts as a tunable parameter to drive transitions between trivial and Chern insulator.
For the sake of completeness, the competition between these two scenarios is also briefly discussed and
summarized in Fig.~\ref{phasediagram3}.

To proceed, we examine the behavior of AHC accompanied by the topological phase transitions
at zero temperature~\cite{Qi2006PRB,Nandy2024arXiv,Habe2025PRB,Baidya2025arXiv}.
Due to the TRSB and nontrivial Berry curvature for both of two scenarios,
AHC exhibits universal quantization $\sigma_{xy} = Ce^2/\hbar$ with $C = 1$ at $E_f=0$.
This is in well agreement with the TKNN formula~\cite{Thouless1982PRL,Kane2010RMP}.
While $E_f\neq 0$, the magnitude of $\sigma_{xy}$ displays parameter-dependent variations as
collected in Fig.~\ref{AHC_pert} and Fig.~\ref{AHC_Floq} for the perturbative and the dynamic
Floquet scenarios, respectively.
In addition, we demonstrate that the QBCP semimetal is characterized by a nontrivial Euler invariant $\chi=1$
with unique optical properties~\cite{Jankowski2025PRB}.
Once $\mathcal{C}_2\mathcal{T}$ symmetry is broken by either static or Floquet driving,
the system enters a Chern insulator phase featured by $C = \pm 1$, which exhibits pronounced circular
dichroism that encodes the Chern-number sign~\cite{Tran2017SA,Asteria2019NP,Chau2025PRB}. This allows the feasibility of
using optical probes, such as circular dichroism~\cite{Jankowski2025PRB, Chau2025PRB},
optical-weight measurements~\cite{Jankowski2025PRB, Chau2025PRB}, and higher-order nonlinear photoconductivity~\cite{Jankowski2025PRB, Chau2025PRB}, to distinguish between distinct topological phases
as summarized in Table.~\ref{tab_II}.

These results establish a tunable topological phase transition from a QBCP semimetal to Chern insulator
in the 2D QBCP materials. The transition yields direct experimental signatures, notably the quantized AHC in
the linear-response regime~\cite{Karplus1954PR,Streda2010PRB,Saha2016arXiv,Qi2006PRB,Nandy2024arXiv,
Habe2025PRB,Baidya2025arXiv} and distinct optical responses associated with different topological
phases~\cite{Jankowski2025PRB, Chau2025PRB,Tran2017SA,Asteria2019NP}.
The unitary equivalence between our model and Bernal bilayer graphene~\cite{Edward2006PRL,Jain2025PRB}
further identifies a concrete material platform for testing these predictions. Looking forward,
future studies on disorder effects~\cite{Jankowski2024PRB} and detailed nonlinear optical
responses~\cite{Edward2006PRL} will be essential to deepen the understanding of topological
control in 2D QBCP and related semimetals.

\section*{ACKNOWLEDGEMENTS}

We thank Wen Liu for the helpful discussions.
J.W. is supported  by Tianjin Natural Science Foundation Project (25JCYBJC01640) and
was partially supported by the National Natural
Science Foundation of China under Grant No. 11504360.

\section*{Data Availability Statement}

Data Availability Statement: No Data associated in
the manuscript.


\appendix

\section{Derivation of the Floquet Hamiltonian}\label{AppB}

We hereby present the derivation of the Floquet Hamiltonian~(\ref{eq16}) in the manintext.
Expanding the full Hamiltonian~(\ref{eq8}) yields~\cite{Polkovnikov2014arXiv}
\begin{eqnarray}
\mathcal{H}_{\mathrm{F}}(\bm{k})=\sum_{n=0}^\infty\mathcal{H}_{\mathrm{F}}^{(n)}(\bm{k}).
\end{eqnarray}
Following the Ref.~\cite{Saha2016arXiv}, we keep the series to the first order,
\begin{eqnarray}
\mathcal{H}_{\mathrm{F}}(\bm{k})\sim \mathcal{H}_{\mathrm{F}}^{(0)}+\mathcal{H}_{\mathrm{F}}^{(1)},\label{B2}
\end{eqnarray}
where the zero-order term is obtained by time-averaging~\cite{Polkovnikov2014arXiv,Saha2016arXiv}:
\begin{eqnarray}
\mathcal{H}_{\mathrm{F}}^{(0)}
&=&
\frac{1}{T}\int_0^T dt~\mathcal{H}(\bm{k},t)\nonumber\\
&=&
\mathcal{H}_0(\bm{k})+t_{I}e^2 A_0^2~\sigma_0+t_x e^2 A_0^2\cos\phi~\sigma_1,\label{Eq_appendix_H0}
\end{eqnarray}
and the first order term is written as
\begin{equation}
\mathcal{H}_{\mathrm{F}}^{(1)}\!=\!\frac{1}{2!iT\hbar}\int_0^T\!dt_1
\int_0^{t_1}\!dt_2\big[\mathcal{H}(\bm{k},t_1), \mathcal{H}(\bm{k},t_2)\big],
\end{equation}
where the middle bracket denotes $[a,b]=ab-ba$.

After several calculations, Eq.~(\ref{B2}) can be reformulated as follows
\begin{eqnarray}
\mathcal{H}(\bm{k},t)
\equiv
F_0(\bm{k},t)\sigma_0+F_1(\bm{k},t)\sigma_1+F_3(\bm{k},t)\sigma_3,
\end{eqnarray}
where
\begin{eqnarray}
F_0(\bm{k},t)&\equiv& t_I\bm{k}^2+H_1(t),\\
F_1(\bm{k},t)&\equiv& 2 t_x k_x k_y+H_2(t),\\
F_3(\bm{k},t)&\equiv& t_z(k_x^2-k_y^2)+H_3(t).
\end{eqnarray}
This leads to
\begin{eqnarray}
&&\big[\mathcal{H}(\bm{k},t_1), \mathcal{H}(\bm{k},t_2)\big]\nonumber\\
&=&\!
2i\!
\left[F_3(\bm{k},t_1)F_1(\bm{k},t_2)\!-\!F_1(\bm{k},t_1)F_3(\bm{k},t_2)\right]\!\sigma_2,
\end{eqnarray}
and the first-order term can be represented by an effective mass,
\begin{eqnarray}
\mathcal{H}_{\mathrm{F}}^{(1)}=m_{\mathrm{eff}}\sigma_2,\label{Eq_appendix_H1}
\end{eqnarray}
where the effective mass is designated as
\begin{widetext}
\begin{eqnarray}
m_{\mathrm{eff}}&=&\frac{eA_0t_xt_z}{4\omega\hbar}\bigg\{8e^2A_0^2
\big[1+\cos(2\phi)\big]k_y+20eA_0\sin\phi~k_x^2+12eA_0\sin\phi~k_y^2+4eA_0\sin(2\phi)k_xk_y\nonumber\\
&&-16 e^2A_0^2\cos\phi~k_x+16(\cos\phi~k_x-k_y)\bm{k}^2+e^3 A_0^3\big[3\sin\phi+\sin(3\phi)\big]\bigg\}.
\end{eqnarray}
\end{widetext}
After combining $\mathcal{H}_{\mathrm{F}}^0$~(\ref{Eq_appendix_H0}) and $\mathcal{H}_{\mathrm{F}}^1$~(\ref{Eq_appendix_H1}),
the final effective Floquet Hamiltonian is then left with
\begin{eqnarray}
\mathcal{H}_{\mathrm{F}}(\bm{k})=t_I(e^2A_0^2+\bm{k}^2)\sigma_0+\bm{d}(\bm{k})\cdot\bm{\sigma},
\end{eqnarray}
where the vector $\bm{d}$ is denominated as
\begin{equation}
\bm{d}(\bm{k})\!=\!(t_xe^2A_0^2\cos\phi+2t_xk_xk_y,m_{\mathrm{eff}},t_z(k_x^2-k_y^2)).
\end{equation}



\begin{thebibliography}{99}

\bibitem{Haldane1988PRL}
F. D. M. Haldane,
\href{https://doi.org/10.1103/PhysRevLett.61.2015}
{Phys. Rev. Lett. {\bf 61}, 2015 (1988)}.

\bibitem{Kane2005PRL}
C. L. Kane and E. J. Mele,
\href{https://doi.org/10.1103/PhysRevLett.95.146802}
{Phys. Rev. Lett. {\bf 95}, 146802 (2005)}.

\bibitem{Kane2010RMP}
M. Z. Hasan and C. L. Kane,
\href{https://doi.org/10.1103/RevModPhys.82.3045}
{Rev. Mod. Phys. {\bf 82}, 3045 (2010)}.

\bibitem{Xiao2010RMP}
D. Xiao, M.-C. Chang, and Q. Niu,
\href{https://doi.org/10.1103/RevModPhys.82.1959}
{Rev. Mod. Phys. {\bf 82}, 1959 (2010)}.

\bibitem{Zhang2011RMP}
X.-L. Qi and S.-C. Zhang,
\href{https://doi.org/10.1103/RevModPhys.83.1057}
{Rev. Mod. Phys. {\bf 83}, 1057 (2011)}.

\bibitem{Feng2012SCPMA}
W. X. Feng and Y. G. Yao,
\href{https://doi.org/10.1007/s11433-012-4929-9}
{Sci. China Phys. Mech. Astron. {\bf 55}, 2199¨C2212 (2012)}.

\bibitem{Chiu2016RMP}
C.-K. Chiu, J. C. Y. Teo, A. P. Schnyder, and S. Ryu,
\href{https://doi.org/10.1103/RevModPhys.88.035005}
{Rev. Mod. Phys. {\bf 88}, 035005 (2016)}.

\bibitem{Yan2012RPP}
B. Yan and S.-C. Zhang,
\href{https://dx.doi.org/10.1088/0034-4885/75/9/096501}
{Rep. Prog. Phys. {\bf 75} 096501 (2012)}.

\bibitem{Ando2013JPSJ}
Yoichi Ando,
\href{https://journals.jps.jp/doi/10.7566/JPSJ.82.102001}
{J. Phys. Soc. Jpn. {\bf 82}, 102001 (2013)}.

\bibitem{Bardarson2013RPP}
J. H. Bardarson and J. E. Moore,
\href{https://iopscience.iop.org/article/10.1088/0034-4885/76/5/056501}
{Rep. Prog. Phys. {\bf 76} 056501 (2013)}.

\bibitem{Beenakker2013ARCMP}
C. W. J. Beenakker,
\href{https://doi.org/10.1146/annurev-conmatphys-030212-184337}
{Annu. Rev. Condens. Matter Phys. {\bf 4}, 113 (2013)}.

\bibitem{Zhang2013RSSRRL}
H. Zhang and S.-C. Zhang,
\href{https://doi.org/10.1002/pssr.201206414}
{Phys. Status Solidi RRL {\bf 7}, 72 (2013)}.

\bibitem{Bansil2016RMP}
A. Bansil, H. Lin, and T. Das,
\href{https://doi.org/10.1103/RevModPhys.88.021004}
{Rev. Mod. Phys. {\bf 88}, 021004 (2016)}.

\bibitem{Lapano2020PRM}
J. Lapano, \emph{et al},
\href{https://doi.org/10.1103/PhysRevMaterials.4.111201}
{Phys. Rev. Materials {\bf 4}, 111201(R) (2020)}.

\bibitem{Bernevig2006SCI}
B. A. Bernevig, T. L. Hughes, and S. C. Zhang,
\href{https://www.science.org/doi/10.1126/science.1133734}
{Science {\bf 314}, 1757 (2006)}.

\bibitem{Thouless1982PRL}
D. J. Thouless, M. Kohmoto, M. P. Nightingale, and M. den Nijs,
\href{https://doi.org/10.1103/PhysRevLett.49.405}
{Phys. Rev. Lett. {\bf 49}, 405 (1982)}.

\bibitem{Chern2020PRL}
R. Chen, C.-Z. Chen, J.-H. Gao, B. Zhou, and D.-H. Xu,
\href{https://doi.org/10.1103/PhysRevLett.124.036803}
{Phys. Rev. Lett. {\bf 124}, 036803 (2020)}.


\bibitem{Chen2021PRX}
Z.-G. Chen, W. Zhu, Y. Tan, L. Wang, and G. Ma,
\href{https://doi.org/10.1103/PhysRevX.11.011016}
{Phys. Rev. X {\bf 11}, 011016 (2021)}.

\bibitem{Bai2025ACS}
Y. Bai, X. Zou, Z. Chen, R. Li, Bo Yuan, Y. Dai, B. Huang, and C. Niu,
\href{https://pubs.acs.org/doi/10.1021/acsnano.5c00323}
{ACS Nano, 19, 9, 9265¨C9272 (2025)}.

\bibitem{Xiang2023NatCommun}
Z.-C. Xiang \emph{et al}.,
\href{https://doi.org/10.1038/s41467-023-41230-9}
{Nat Commun {\bf 14}, 5433 (2023)}.

\bibitem{Lu2025PRL}
H. Lu, H.-Q. Wu, B.-B. Chen, and Z. Y. Meng,
\href{https://doi.org/10.1103/PhysRevLett.134.076601}
{Phys. Rev. Lett. {\bf 134}, 076601 (2025)}.

\bibitem{Wang2025arXiv}
B, Wang, J. Yu, P. Sharma, and C.-C. Liu,
\href{https://doi.org/10.48550/arXiv.2504.11177}
{arXiv:2504.11177 [cond-mat.mes-hall] (2025)}.

\bibitem{He2025arXiv}
Y. He, S.H. Simon, and S.A. Parameswaran,
\href{https://doi.org/10.48550/arXiv.2505.06354}
{arXiv:2505.06354 [cond-mat.str-el] (2025)}.

\bibitem{Emanuel2025arXiv}
P. Emanuel, A. Keselman, and Y. Oreg,
\href{https://doi.org/10.1103/9hpw-kz7g}
{Phys. Rev. B {\bf 112}, 235133 (2025)}.

\bibitem{Lin2025arXiv}
Z. Lin, W. Yang, H. Lu, D. Zhai, and W. Yao,
\href{https://doi.org/10.1016/j.newton.2025.100339}
{ Newton {\bf 2}, 100339 (2025)}.

\bibitem{Chang2013SCI}
C.-Z. Chang \emph{et al}.,
\href{https://doi.org/10.1126/science.1234414}
{Science {\bf 340}, 167-170 (2013)}.

\bibitem{Huang2024JPCM}
Y. Huang, Y. Fu, P. Zhang, K. L. Wang, and Q. L. He,
\href{https://doi.org/10.1088/1361-648X/ad550a}
{J. Phys.: Condens. Matter {\bf 36} 37LT01 (2024)}.

\bibitem{Zhang2023PRB}
X. Zhang, G. Pan, B.-B. Chen, H. Li, K. Sun, and Z. Y. Meng,
\href{https://doi.org/10.1103/PhysRevB.107.L241105}
{Phys. Rev. B {\bf 107}, L241105 (2023)}.

\bibitem{Ji2024Nature}
Z. Ji, H. Park, M. E. Barber, C. Hu, K. Watanabe, T. Taniguchi, J.-H. Chu, X. Xu, and Z.-X. Shen,
\href{https://doi.org/10.1038/s41586-024-08092-7}
{Nature {\bf 635}, 578¨C583 (2024)}.

\bibitem{Lei2024CPL}
L.-X. Lei \emph{et al},
\href{https://doi.org/10.1088/0256-307X/41/9/090301}
{Chinese Phys. Lett. {\bf 41} 090301 (2024)}.

\bibitem{Castro2009RMP}
A. H. Castro Neto, F. Guinea, N. M. R. Peres, K. S. Novoselov, and A. K. Geim,
\href{https://doi.org/10.1103/RevModPhys.81.109}
{Rev. Mod. Phys. {\bf 81}, 109 (2009)}.

\bibitem{Lu2010PRB}
H.-Z. Lu, W.-Y. Shan, W. Yao, Q. Niu, and S.-Q. Shen,
\href{https://doi.org/10.1103/PhysRevB.81.115407}
{Phys. Rev. B {\bf 81}, 115407 (2010)}.

\bibitem{Shen2011SPIN}
S.-Q. Shen, W.-Y. Shan, and H.-Z. Lu,
\href{https://doi.org/10.1142/S2010324711000057}
{SPIN 01:01, 33-44 (2011)}.

\bibitem{Taguchi2020PRB}
K. Taguchi, D. Oshima, Y. Yamaguchi, T. Hashimoto, Y. Tanaka, and M. Sato,
\href{https://doi.org/10.1103/PhysRevB.101.235201}
{Phys. Rev. B {\bf 101}, 235201 (2020)}.

\bibitem{Vargiamidis2022arXiv}
V. Vargiamidis, P. Vasilopoulos, and N. Neophytou,
\href{https://doi.org/10.48550/arXiv.2212.10667}
{arXiv:2212.10667 [cond-mat.mes-hall] (2022)}.

\bibitem{Lu2022PRB}
H. Lu, S. Sur, S.-S. Gong, and D. N. Sheng,
\href{https://doi.org/10.1103/PhysRevB.106.205105}
{Phys. Rev. B {\bf 106}, 205105 (2022)}.

\bibitem{Wang2023PRB}
Z.-M. Wang, R. Wang, J.-H. Sun, T.-Y. Chen, and D.-H. Xu,
\href{https://doi.org/10.1103/PhysRevB.107.L121407}
{Phys. Rev. B {\bf 107}, L121407 (2023)}.

\bibitem{Mo2024arXiv}
Y. Mo, X. Wang, Z.-Y. Zhuang, and Z. Yan,
\href{https://doi.org/10.1103/PhysRevB.111.L140504}
{Phys. Rev. B {\bf 111}, L140504 (2025)}.

\bibitem{Huang2015PRB}
H. Huang, Z. Liu, H. Zhang, W. Duan, and D. Vanderbilt,
\href{https://doi.org/10.1103/PhysRevB.92.161115}
{Phys. Rev. B {\bf 92}, 161115(R) (2015)}.

\bibitem{Saha2016arXiv}
K. Saha,
\href{https://doi.org/10.1103/PhysRevB.94.081103}
{Phys. Rev. B {\bf 94}, 081103(R) (2016)}.

\bibitem{Mondal2022PRB}
S. Mondal and S. Basu,
\href{https://doi.org/10.1103/PhysRevB.105.235441}
{Phys. Rev. B {\bf 105}, 235441 (2022)}.

\bibitem{Chen2018PRB}
Q. Chen, L. Du, and G. A. Fiete,
\href{https://doi.org/10.1103/PhysRevB.97.035422}
{Phys. Rev. B {\bf 97}, 035422 (2018)}.

\bibitem{Zhang2005Nature}
Y. Zhang, Y.-W. Tan, H. L. Stormer, and P. Kim,
\href{https://www.nature.com/articles/nature04235}
{Nature {\bf 438}, 201¨C204 (2005)}.

\bibitem{Chong2008PRB}
Y. D. Chong, X.-G. Wen, and M. Soljacic,
\href{https://doi.org/10.1103/PhysRevB.77.235125}
{Phys. Rev. B {\bf 77}, 235125 (2008)}.

\bibitem{Sun2008PRB}
K. Sun and E. Fradkin,
\href{https://doi.org/10.1103/PhysRevB.78.245122}
{Phys. Rev. B {\bf 78}, 245122 (2008)}.

\bibitem{Sun2009PRL}
K. Sun, H. Yao, E. Fradkin, and S. A. Kivelson,
\href{https://doi.org/10.1103/PhysRevLett.103.046811}
{Phys. Rev. Lett. {\bf 103}, 046811 (2009)}.

\bibitem{Vafek2014PRB}
J. M. Murray and O. Vafek,
\href{https://doi.org/10.1103/PhysRevB.89.201110}
{Phys. Rev. B {\bf 89}, 201110(R) (2014)}.

\bibitem{Yao2022PRR}
M.-R. Li, A.-L. He, and H. Yao,
\href{https://doi.org/10.1103/PhysRevResearch.4.043151}
{Phys. Rev. Research {\bf 4}, 043151 (2022)}.

\bibitem{Wang2024AP}
Y.-S. Fu and J. Wang,
\href{https://doi.org/10.1016/j.aop.2024.169811}
{Annals of Physics {\bf 470}, 169811 (2024)}.

\bibitem{Mandal2019CMP}
I. Mandal and S. Gemsheim,
\href{https://doi.org/10.5488/CMP.22.13701}{Condens. Matter Phys., {\bf 22},1, 13701 (2019)}.

\bibitem{Bera2021JPCM}
S. Bera and I. Mandal,
\href{https://iopscience.iop.org/article/10.1088/1361-648X/ac020a}{J. Phys.: Condens. Matter {bf 33} 295502 (2021)}.

\bibitem{Lu2024PRB}
H. Lu, K. Sun, Z. Y. Meng, and B.-B. Chen,
\href{https://doi.org/10.1103/PhysRevB.109.L081106}{Phys. Rev. B {\bf 109}, L081106 (2024)}.


\bibitem{Liquito2024PRB}
R. Liquito, M. Goncalves, and E. V. Castro,
\href{https://doi.org/10.1103/PhysRevB.109.174202}
{Phys. Rev. B {\bf 109}, 174202 (2024)}.

\bibitem{Wan2023PRL}
X. Wan, S. Sarkar, S.-Z. Lin, and K. Sun,
\href{https://doi.org/10.1103/PhysRevLett.130.216401}
{Phys. Rev. Lett. {\bf 130}, 216401 (2023)}.

\bibitem{Ji2022PRB}
X. Ji, J. Gao, C. Yue, Z. Wang, H. Wu, X. Dai, and H. Weng,
\href{https://doi.org/10.1103/PhysRevB.106.235103}
{Phys. Rev. B {\bf 106}, 235103 (2022)}.

\bibitem{Sobrosa2024PRB}
N. Sobrosa, M. Goncalves, and E. V. Castro,
\href{https://doi.org/10.1103/PhysRevB.109.184206}
{Phys. Rev. B {\bf 109}, 184206 (2024)}.

\bibitem{Jung2023arXiv}
J. Jung, H. Lim, and B.-J. Yang,
\href{https://doi.org/10.48550/arXiv.2307.12528}
{arXiv:2307.12528 [cond-mat.mes-hall] (2023)}.

\bibitem{Wu2022SCPMA}
D. Wu, Y. Huang, S. Sun, J. Gao, Z. Guo, H. Weng, Z. Fang, K. Jiang, and Z. Wang,
\href{https://doi.org/10.1007/s11433-021-1862-3}
{Sci. China Phys. Mech. Astron. {\bf 65}, 256811 (2022)}.

\bibitem{Wang2017PRB}
J. Wang, C. Ortix, J. van den Brink, and D. V. Efremov,
\href{https://journals.aps.org/prb/abstract/10.1103/PhysRevB.96.201104}
{Phys. Rev. B {\bf 96}, 201104(R) (2017)}.

\bibitem{DZZW2020PRB}
Y. -M. Dong, Y.- H. Zhai, D. -X. Zheng, and J. Wang,
\href{https://doi.org/10.1103/PhysRevB.102.134204}
{Phys. Rev. B {\bf 102}, 134204 (2020).}

\bibitem{Janssen2018PRB}
S. Ray, M. Vojta, and L. Janssen,
\href{https://journals.aps.org/prb/abstract/10.1103/PhysRevB.98.245128}
{Phys. Rev. B {\bf 98}, 245128 (2018)}.

\bibitem{Pan2015SR}
H. Pan, M. Wu, Yl Liu, and S. A. Yang,
\href{https://doi.org/10.1038/srep14639}{Sci. Rep. {\bf 5}, 14639 (2015)}.

\bibitem{Oka2009PRB}
T. Oka and H. Aoki,
\href{https://doi.org/10.1103/PhysRevB.79.081406}{Phys. Rev. B {\bf 79}, 081406(R) (2009)}.

\bibitem{Rudner2020NRP}
M. S. Rudner and N. H. Lindner,
\href{https://doi.org/10.1038/s42254-020-0170-z}{Nat. Rev. Phys. {\bf 2}, 229¨C244 (2020)}.

\bibitem{Liu2025arXiv}
P. Liu, C. Cui, L. Li, R. Li, D.-H. Xu, and Z.-M. Yu,
\href{https://doi.org/10.48550/arXiv.2503.19614}{arXiv:2503.19614 [cond-mat.str-el] (2025)}.

\bibitem{Fu2025arXiv}
P.-H. Fu, S. Mondal, J.-F. Liu, Y. Tanaka, and J. Cayao,
\href{https://doi.org/10.48550/arXiv.2505.20205}{arXiv:2505.20205 [cond-mat.supr-con] (2025)}.

\bibitem{Yokoyama2025arXiv}
T. Yokoyama,
\href{https://doi.org/10.1103/4tng-rhc4}
{Phys. Rev. B {\bf 112}, 024512 (2025)}.

\bibitem{Jankowski2025PRB}
W. J. Jankowski, A. S. Morris, A. Bouhon, F. NurUnal, and R.-J. Slager,
\href{https://doi.org/10.1103/PhysRevB.111.L081103}
{Phys. Rev. B {\bf 111}, L081103 (2025)}.

\bibitem{Chau2025PRB}
C. W. Chau, W. J. Jankowski, and R.-J. Slager,
\href{https://doi.org/10.1103/b9ll-927t}
{Phys. Rev. B {\bf 112}, 064512 (2025)}.

\bibitem{Tran2017SA}
D. T. Tran, A. Dauphin, A. G. Grushin, P. Zoller, and N.
Goldman,
\href{https://www.science.org/doi/10.1126/sciadv.1701207}
{Sci. Adv. {\bf 3}, e1701207 (2017)}.

\bibitem{Asteria2019NP}
L. Asteria \emph{et al}.,
\href{https://www.nature.com/articles/s41567-019-0417-8}
{Nat. Phys. {\bf 15}, 449 (2019)}.

\bibitem{Polkovnikov2014arXiv}
M. Bukov, L. D'Alessio, and A. Polkovnikov,
\href{https://doi.org/10.1080/00018732.2015.1055918}
{Advances in Physics, {\bf 64}, 2, 139-226 (2015)}.

\bibitem{Benhaida2024arXiv}
O. Benhaida, E. H. Saidi, L. B. Drissi, and R. Ahl Laamara,
\href{https://doi.org/10.48550/arXiv.2412.17763}
{arXiv:2412.17763 [cond-mat.mes-hall] (2024)}.

\bibitem{Kibis2024arXiv}
O. V. Kibis, M. V. Boev, I. V. Iorsh, and V. M. Kovalev,
\href{https://doi.org/10.1088/1361-648X/ad88c5}
{J. Phys.: Condens. Matter {\bf 37}, 035503 (2025)}.

\bibitem{Asgari2025arXiv}
S. Sajad Dabiri and Reza Asgari,
\href{https://doi.org/10.48550/arXiv.2503.12620}
{arXiv:2503.12620 [cond-mat.mtrl-sci] (2025)}.

\bibitem{Wang2021NPB}
Y.-H. Zhai and J. Wang,
\href{https://doi.org/10.1016/j.nuclphysb.2021.115371}
{Nucl. Phys. B {\bf 966} (2021) 115371}.

\bibitem{Zhou2007NM}
S. Y. Zhou, G.-H. Gweon, A. V. Fedorov, P. N. First, W. A. de Heer,
D.-H. Lee, F. Guinea, A. H. Castro Neto, and A. Lanzara,
\href{https://doi.org/10.1038/nmat2003}
{Nature Mater {\bf 6}, 770¨C775 (2007)}.

\bibitem{Bao2021PRL}
C.-H. Bao, H.-Y. Zhang, T. Zhang, X. Wu, L.-P. Luo, S.-H. Zhou, Q. Li, Y.-H. Hou, W. Yao \emph{et al}.
\href{https://doi.org/10.1103/PhysRevLett.126.206804}
{Phys. Rev. Lett. {\bf 126}, 206804 (2021)}.

\bibitem{Wang2025PRB}
Z. Wang, J.-S. Yan, K.-X. Chen, and S.-S. Lyu,
\href{https://doi.org/10.1103/PhysRevB.111.085425}
{Phys. Rev. B {\bf 111}, 085425 (2025)}.

\bibitem{Wang2013PRL}
Z. F. Wang, Z. Liu, and F. Liu,
\href{https://doi.org/10.1103/PhysRevLett.110.196801}
{Phys. Rev. Lett. {\bf 110}, 196801 (2013)}.

\bibitem{Wu2016PRL}
H.-Q. Wu, Y.-Y. He, C. Fang, Z. Y. Meng, and Z.-Y. Lu,
\href{https://doi.org/10.1103/PhysRevLett.117.066403}
{Phys. Rev. Lett. {\bf 117}, 066403 (2016)}.

\bibitem{Ma2024arXiv}
F. Ma, J. Feng, F. Li, Y. Wu, and D. Zhou,
\href{https://doi.org/10.48550/arXiv.2412.00619}
{arXiv:2412.00619 [cond-mat.mtrl-sci] (2024)}.

\bibitem{Mola2025arXiv}
Z. Jalali-Mola and O. Hess,
\href{https://doi.org/10.48550/arXiv.2503.17263}
{arXiv:2503.17263 [physics.optics] (2025)}.

\bibitem{Li2025arXiv}
C. Li, \emph{et al},
\href{https://doi.org/10.48550/arXiv.2505.01767}
{arXiv:2505.01767 [cond-mat.mes-hall] (2025)}.

\bibitem{Pires2024arXiv}
P. G. de Oliveira and A. S. T. Pires,
\href{https://doi.org/10.48550/arXiv.2407.20296}
{arXiv:2407.20296 [cond-mat.mes-hall] (2024)}.

\bibitem{Sin2024arXiv}
H.-B. Kim, T. Yuk, and S.-J. Sin,
\href{https://doi.org/10.48550/arXiv.2407.21098}
{arXiv:2407.21098 [cond-mat.mes-hall] (2024)}.

\bibitem{Singh2024arXiv}
A. Mukherjee and B. Singh,
\href{https://doi.org/10.48550/arXiv.2410.04515}
{arXiv:2410.04515 [cond-mat.mes-hall] (2024)}.

\bibitem{Villegas2024arXiv}
J. O. A. Biscocho and K. H. A. Villegas,
\href{https://doi.org/10.48550/arXiv.2410.14213}
{arXiv:2410.14213 [cond-mat.supr-con] (2024)}.

\bibitem{Qi2006PRB}
X.-L. Qi, Y.-S. Wu, and S.-C. Zhang,
\href{https://doi.org/10.1103/PhysRevB.74.085308}
{Phys. Rev. B {\bf 74}, 085308 (2006)}.

\bibitem{Nandy2024arXiv}
S. Pradhan, K. Samanta, and A. K. Nandy,
\href{https://doi.org/10.48550/arXiv.2412.02324}
{arXiv:2412.02324 [cond-mat.mtrl-sci] (2024)}.

\bibitem{Habe2025PRB}
T. Habe,
\href{https://doi.org/10.1103/PhysRevB.111.035303}
{Phys. Rev. B {\bf 111}, 035303 (2025)}.

\bibitem{Karplus1954PR}
R. Karplus and J. M. Luttinger,
\href{https://doi.org/10.1103/PhysRev.95.1154}
{Phys. Rev. {\bf 95}, 1154 (1954)}.

\bibitem{Streda2010PRB}
P. St\u{r}eda,
\href{https://doi.org/10.1103/PhysRevB.82.045115}{Phys. Rev. B {\bf 82}, 045115 (2010)}.

\bibitem{Baidya2025arXiv}
S. Thakur and S. Baidya,
\href{https://doi.org/10.48550/arXiv.2501.11025}
{arXiv:2501.11025 [cond-mat.mtrl-sci] (2025)}.

\bibitem{Hall1879AJOM}
E. H. Hall,
\href{https://doi.org/10.2307/2369245}{American Journal of Mathematics, 2(3): 287-292 (1879)}.

\bibitem{Klitzing1986RMP}
K. v. Klitzing,
\href{https://doi.org/10.1103/RevModPhys.58.519}{Rev. Mod. Phys. {\bf 58}, 519 (1986)}.



\bibitem{Jain2025PRB}
A. Jain, W. J. Jankowski, and R.-J. Slager,
\href{https://doi.org/10.1103/5gmg-q1z5}
{Phys. Rev. B {\bf 111}, 235149 (2025)}.

\bibitem{Bouhon2020NP}
A. Bouhon, Q. Wu, R.-J. Slager, H. Weng, O. V. Yazyev, and T. Bzdusek,
\href{https://www.nature.com/articles/s41567-020-0967-9}
{Nat. Phys. {\bf 16}, 1137 (2020)}.

\bibitem{Bouhon2022arXiv}
A. Bouhon and R.-J. Slager,
\href{https://doi.org/10.48550/arXiv.2203.16741}
{arXiv: 2203.16741 [cond-mat.mes-hall] (2022)}.

\bibitem{Edward2006PRL}
E. McCann and V. I. Falko,
\href{https://doi.org/10.1103/PhysRevLett.96.086805}
{Phys. Rev. Lett. {\bf 96}, 086805 (2006)}.

\bibitem{Jankowski2024PRB}
W. J. Jankowski, M. Noormandipour, A. Bouhon, and R.-J. Slager,
\href{https://doi.org/10.1103/PhysRevB.110.064202}
{Phys. Rev. B {\bf 110}, 064202 (2024)}.


\end{thebibliography}
\end{document}